\documentclass[prd,aps,showpacs,epsf,floats]{revtex4}%
\usepackage{amssymb}
\usepackage{amsfonts}
\usepackage{amsmath}
\usepackage{graphicx}%
\providecommand{\U}[1]{\protect\rule{.1in}{.1in}}
\begin{document}
\title{\textbf{A simple comparative analysis of exact and approximate quantum error
correction}}
\author{Carlo Cafaro$^{1,2}$ and Peter van Loock$^{2}$}
\affiliation{$^{1}$Max-Planck Institute for the Science of Light, 91058 Erlangen, Germany }
\affiliation{$^{2}$Institute of Physics, Johannes-Gutenberg University Mainz, 55128 Mainz, Germany}

\begin{abstract}
We present a comparative analysis of exact and approximate quantum error
correction by means of simple unabridged analytical computations. For the sake
of clarity, using primitive quantum codes, we study the exact and approximate
error correction of the two simplest unital (Pauli errors) and nonunital
(non-Pauli errors) noise models, respectively. The similarities and
differences between the two scenarios are stressed. In addition, the
performances of quantum codes quantified by means of the entanglement fidelity
for different recovery schemes are taken into consideration in the approximate
case. Finally, the role of self-complementarity in approximate quantum error
correction is briefly addressed.

\end{abstract}

\maketitle

\section{Introduction}

It is known that decoherence is one of the most important obstacles in quantum
information processing, since it causes a quantum computer to lose its quantum
properties destroying its performance advantages over a classical computer.
There are different methods for preserving quantum coherence. One possible
technique exploits redundancy in encoding information. As pointed out in
\cite{dv}, one might think that redundancy cannot be of any use in quantum
computing, since quantum states cannot be cloned \cite{dolly}. However, using
the property of quantum entanglement, Shor and Steane discovered a clever
scheme for exploiting redundancy \cite{shor1, steane}. This scheme is known as
\textit{quantum error correcting codes} (QECCs). For a comprehensive
introduction to QECCs, we refer to \cite{daniel}. Within such scheme,
information is encoded in linear subspaces (codes) of the total Hilbert space
in such a way that errors induced by the interaction with the environment can
be detected and corrected. The QECCs approach may be interpreted as an active
stabilization of a quantum state in which, by monitoring the system and
conditionally carrying on suitable operations, one prevents the loss of
information. In detail, the errors occur on a qubit when its evolution differs
from the ideal one. This happens by interaction of the qubit with an environment.

Among the first and most famous QECCs, there are the Shor nine-qubit code
\cite{shor1}, the Calderbank-Shor-Steane seven-qubit code \cite{robert1,
steane} and the perfect $1$-error correcting five-qubit code \cite{laflamme,
bennett} with transmission rate equal to $1/5$. In general, scientists aim at
searching for new quantum codes capable of combatting very general error
models and correcting for arbitrary errors at unknown positions in the
codeword. These results, in general, have more theoretical than practical
importance, since they assume the existence of a fairly sophisticated quantum
computer which has not been built yet. Fortunately, in many realistic
situations, additional information on possible errors is available. Ideally,
this knowledge should be taken into consideration in order to construct the
simplest code with the highest transmission rate that can have a good chance
to be implemented in a real laboratory. We stress \textit{ideally} since, in
general, it is a difficult problem to design quantum codes for any particular
noise model. For instance, there are quantum systems for which the noise leads
to dephasing errors only or bit-flip errors only. It has been shown that for
such restricted types of decoherence, it is possible to perform error
correction of one arbitrary dephasing/bit-flip error by encoding a single
logical qubit into a minimum of three physical qubits \cite{sam}. Furthermore,
uncovering efficient codes for restricted error models may be important for
proof of principle demonstrations of quantum error correction \cite{markus}.
Consider an error model where the position of the erroneous qubits is known.
Such errors at known positions are denoted as \emph{erasures }\cite{markus}%
\emph{. }A $t$-error correcting code is a $2t$-erasure correcting code. Also,
while $1/5$ is the highest rate for a $1$-error correcting code, four qubits
are sufficient for a code to correct one arbitrary erasure. Omitting suitable
normalization factors, the perfect four-qubit code for the correction of one
erasure reads \cite{markus},%
\begin{equation}
\left\vert 0_{L}\right\rangle \overset{\text{def}}{=}\left\vert
0000\right\rangle +\left\vert 1111\right\rangle \text{ and, }\left\vert
1_{L}\right\rangle \overset{\text{def}}{=}\left\vert 1001\right\rangle
+\left\vert 0110\right\rangle \text{. } \label{markus}%
\end{equation}
In general, when no knowledge on the noise model is assumed, the errors to be
corrected are completely random. This scenario may lead to the error
correction of Pauli-type errors $X$, $Y$, $Z$ (with $X\overset{\text{def}}%
{=}\sigma_{x}$, $Y\overset{\text{def}}{=}\sigma_{y}$, $Z\overset{\text{def}%
}{=}\sigma_{z})$ that occur with equal probability $p_{X}=p_{Y}=p_{Z}=\frac
{p}{3}$ (symmetric depolarizing channel, \cite{nielsenbook}). However, if
further information about an error process is available, more efficient codes
can be designed as pointed out earlier. As a matter of fact, in many physical
systems, the types of noise are likely to be unbalanced between amplitude
($X$-type) and phase ($Z$-type) errors. These are asymmetric error models
which are still described by Kraus operators that are (unitary) Pauli matrices
(Pauli Kraus operators). However, there are types of noise models seen in
realistic settings that are not described by Pauli Kraus operators and, for
these cases as well, the task of constructing good error correcting codes is
very challenging. The amplitude damping (AD) channel is the simplest nonunital
channel whose Kraus operators cannot be described by (unitary) Pauli
operations \cite{nielsenbook}. The two Kraus operators for AD noise are given
by $A_{0}\overset{\text{def}}{=}I-\mathcal{O}\left(  \gamma\right)  $ and
$A_{1}\overset{\text{def}}{=}\sqrt{\gamma}\left\vert 0\right\rangle
\left\langle 1\right\vert $ where $\gamma$ denotes the damping rate (or,
damping probability parameter). As we may observe, there is no simple way of
reducing $A_{1}$ to one Pauli error operator since $\left\vert 0\right\rangle
\left\langle 1\right\vert $ is not normal. Observe that $A_{1}^{\dagger
}\propto\sigma_{x}-i\sigma_{y}$, therefore the linear span of $A_{1}$ and
$A_{1}^{\dagger}$ equals the linear span of $\sigma_{x}$ and $\sigma_{y}$. If
the quantum system interacts with an environment at finite temperature, the
Kraus operator $A_{1}^{\dagger}$ will appear in the noise model (as stressed
in \cite{beny}, the error space to be corrected is a subspace of that spanned
by the interaction operators, selected by the initial state of the
environment) \cite{nielsenbook}. Therefore, if a code is capable of correcting
$t$ $\sigma_{x}$- and $t$ $\sigma_{y}$-errors, it can also correct $t$ $A_{1}$
and $t$ $A_{1}^{\dagger}$ errors. For the AD channel, we only need to deal
with the error $A_{1}$ but not with $A_{1}^{\dagger}$. For such a reason,
requiring to be capable of correcting both $\sigma_{x}$- and $\sigma_{y}%
$-errors is a less efficient way for constructing quantum codes for the AD
channel. The first quantum code correcting single-AD errors was a $\left[
\left[  4,1\right]  \right]  $ code presented by Leung et \textit{al}. in
\cite{leung}. Omitting proper normalization factors, the Leung et \textit{al}.
$\left[  \left[  4,1\right]  \right]  $ code reads,%
\begin{equation}
\left\vert 0_{L}\right\rangle \overset{\text{def}}{=}\left\vert
0000\right\rangle +\left\vert 1111\right\rangle \text{ and, }\left\vert
1_{L}\right\rangle \overset{\text{def}}{=}\left\vert 0011\right\rangle
+\left\vert 1100\right\rangle \text{. } \label{leung}%
\end{equation}
The four-qubit code spanned by the codewords $\left\vert 0_{L}\right\rangle $
and $\left\vert 1_{L}\right\rangle $ in Eq. (\ref{leung}) represents a
departure from standard QECCs that seek to \emph{perfectly} correct up to $t$
arbitrary errors on the system. The key-point advanced in \cite{leung} is that
exact correctability is too strong a restriction. Relaxing the Knill-Laflamme
QEC conditions (KL-conditions) \cite{knill-laflamme} in such a manner that
they are only \emph{approximately} satisfied and allowing for a negligible
error in the recovery scheme, better codes with higher transmission rates can
be uncovered. The adaptation of the code to the noise model, an idea remarked
later also in \cite{fletcher1}, is a crucial factor behind the success of the
Leung et \textit{al}. four-qubit code. Following the lead of \cite{leung},
other works concerning the error correction of amplitude damping errors have
appeared into the literature \cite{fletcher1, lang, shor2, duan, fletcher2,
ng}. In \cite{lang, shor2}, it is emphasized that the concept of
self-complementarity is crucial for error correcting amplitude damping errors,
although the sel-complementarity of the Leung et \textit{al}. code is not
specifically mentioned. In \cite{fletcher1}, the performance of various
quantum error correcting codes for AD errors are numerically analyzed. In
particular, a numerical analysis of the performance (quantified by means of
Schumacher's entanglement fidelity, \cite{benji} ) of the four-qubit code for
two recovery schemes can be found. The recovery schemes employed are the
so-called code projected and the optimal channel adapted recovery schemes. The
latter scheme was computed via semidefinite programming methods in
\cite{fletcher2}. However, no explicit analytical investigation (similar, for
instance, to the investigations presented in \cite{cafaro1} and \cite{cafaro2}
\ for depolarizing and Weyl unitary errors, respectively) is available.
Furthermore, as pointed out in \cite{ng}, numerically computed recovery maps
are difficult to describe and understand analytically.

Inspired by \cite{lang, shor2}, we uncover that among the possible $28$-pairs
of orthonormal self-complementary codewords in $\mathcal{H}_{2}^{4}$, only
three pairs are indeed good (locally permutation equivalent, \cite{markus2})
single AD-error correcting codes. In particular, two of these pairs define the
perfect $1$-erasure correcting code in Eq. (\ref{markus})\ and the well-known
four-qubit Leung et \textit{al}. code in Eq. (\ref{leung}). Moreover,
motivated by the numerical analysis presented in \cite{fletcher1}, we provide
a fully analytical investigation of the performances, quantified in terms of
Schumacher's entanglement fidelity, of the Leung et \textit{al}. four-qubit
code. For the four-qubit code, the performance is evaluated for three
different recovery schemes: the standard QEC recovery operation, the
code-projected recovery operation and, finally, an analytically-optimized
Fletcher's-type channel-adapted recovery operation \cite{fletcher1}.

The layout of this article is as follows. In Section II, we describe necessary
conditions for approximate quantum error correction together with necessary
and sufficient conditions for exact quantum error correction. In Section III,
we present a detailed study for the exact error correction of the bit-flip
(or, similarly, phase-flip) noise errors by means of the three-qubit
repetition code. We also present an analytical investigation of amplitude
errors by means of the Leung et \textit{al}. four-qubit code. The performance
of each code is quantified by means of the entanglement fidelity. In
particular, we compare three different recovery schemes in the approximate
case. The concluding remarks appear in Section IV. A number of appendices with
technical details of calculations are also provided.

\section{From exact to approximate quantum error correction conditions}

The very first \emph{sufficient conditions} for approximate quantum error
correction were introduced by Leung et \textit{al}. in \cite{leung}. They
showed that quantum codes can be effective in the error correction procedure
even though they violated the standard KL-conditions. However, these
violations characterized by small deviations from the standard
error-correction conditions are allowed provided that they do not affect the
desired fidelity order.

\subsection{Exact quantum error correction conditions}

For the sake of reasoning, let us consider a quantum stabilizer code
$\mathcal{C}$ with code parameters $\left[  \left[  n,k,d\right]  \right]  $
encoding $k$-logical qubits in the Hilbert space $\mathcal{H}_{2}^{k}$ into
$n$-physical qubits in the Hilbert space $\mathcal{H}_{2}^{n}$ and distance
$d$. Assume that the noise model after the encoding procedure is
$\Lambda\left(  \rho\right)  $ and can be described by an operator-sum
representation,%
\begin{equation}
\Lambda\left(  \rho\right)  \overset{\text{def}}{=}\sum_{k\in\mathcal{K}}%
A_{k}\rho A_{k}^{\dagger}\text{,}%
\end{equation}
where $\mathcal{K}$ is the index set of all the enlarged Kraus operators
$A_{k}$ that appear in the sum. The noise channel\textbf{ }$\Lambda$\textbf{
}is a CPTP (completely\textbf{ }positive and trace preserving) map. The
codespace of $\mathcal{C}$ is a $k$-dimensional subspace of $\mathcal{H}%
_{2}^{n}$ where some error operators that characterize the error model
$\Lambda$ being considered can be reversed. Denote with $\mathcal{A}%
_{\text{reversible}}\subset$ $\mathcal{A}\overset{\text{def}}{=}\left\{
A_{k}\right\}  $ with $k\in\mathcal{K}$ the set of reversible enlarged errors
$A_{k}$ on $\mathcal{C}$ such that $\mathcal{K}_{\text{reversible}}%
\overset{\text{def}}{=}\left\{  k:A_{k}\in\mathcal{A}_{\text{reversible}%
}\right\}  $ is the index set of $\mathcal{A}_{\text{reversible}}$. Therefore,
the noise model $\Lambda^{\prime}\left(  \rho\right)  $ given by,%
\begin{equation}
\Lambda^{\prime}\left(  \rho\right)  \overset{\text{def}}{=}\sum
_{k\in\mathcal{K}_{\text{reversible}}}A_{k}\rho A_{k}^{\dagger}\text{,}
\label{g1}%
\end{equation}
is reversible on $\mathcal{C}\subset\mathcal{H}_{2}^{n}$. The noise
channel\textbf{ }$\Lambda^{\prime}$\textbf{ }denotes a CP but non-TP map. The
enlarged error operators $A_{k}$ in $\mathcal{A}_{\text{reversible}}$ satisfy
the standard error correction conditions \cite{nielsenbook},%
\begin{equation}
P_{\mathcal{C}}A_{l}^{\dagger}A_{m}P_{\mathcal{C}}=\alpha_{lm}P_{\mathcal{C}%
}\text{,} \label{vai}%
\end{equation}
for any $l$, $m\in\mathcal{K}_{\text{reversible}}$, $P_{\mathcal{C}}$ denotes
the projector on the codespace and $\alpha_{lm}$ are entries of a positive
Hermitian matrix. Furthermore, a subset of error operators $A_{k}$ in
$\mathcal{A}_{\text{reversible}}$ is detectable if it satisfies the following
detectability conditions,
\begin{equation}
P_{\mathcal{C}}A_{k}P_{\mathcal{C}}=\lambda_{A_{k}}P_{\mathcal{C}}\text{,}%
\end{equation}
where $\lambda_{A_{k}}$ denotes a proportionality constant between
$P_{\mathcal{C}}A_{k}P_{\mathcal{C}}$ and $P_{\mathcal{C}}$. The fulfillment
of Eq. (\ref{vai}) for some subset of enlarged error operators $A_{k}$ that
characterize the operator sum representation of the noise model $\Lambda$
implies that there exists an new operator-sum decomposition of $\Lambda$ such
that $\Lambda^{\prime}\left(  \rho\right)  $ in Eq. (\ref{g1}) becomes,%
\begin{equation}
\Lambda^{\prime}\left(  \rho\right)  \overset{\text{def}}{=}\sum
_{k\in\mathcal{K}_{\text{reversible}}^{\prime}}A_{k}^{\prime}\rho
A_{k}^{\prime\dagger}\text{,}%
\end{equation}
where (\ref{vai}) is replaced by%
\begin{equation}
P_{\mathcal{C}}A_{l}^{\prime\dagger}A_{m}^{\prime}P_{\mathcal{C}}=p_{m}%
\delta_{lm}P_{\mathcal{C}}\text{,} \label{vai2}%
\end{equation}
for any $l$, $m\in\mathcal{K}_{\text{reversible}}^{\prime}$ with the error
detection probabilities $p_{m}$ non-negative $c$-numbers. We remark that Eq.
(\ref{vai2}) is equivalent to the usual (exact) orthogonality and
non-deformation conditions for a nondegenerate code,%
\begin{equation}
\left\langle i_{L}\left\vert A_{l}^{\dagger}A_{m}\right\vert j_{L}%
\right\rangle =\delta_{ij}\delta_{lm}p_{m} \label{cacchio}%
\end{equation}
for any $i$, $j$ labelling the logical states and $l$, $m\in\mathcal{K}%
_{\text{reversible}}$.

Observe that for any linear operator $A_{k}^{\prime}$ on a vector space $V$
there exists a unitary $U_{k}$ and a positive operator $J\overset{\text{def}%
}{=}\sqrt{A_{k}^{\prime\dagger}A_{k}^{\prime}}$ such that \cite{nielsenbook},%
\begin{equation}
A_{k}^{\prime}=U_{k}J=U_{k}\sqrt{A_{k}^{\prime\dagger}A_{k}^{\prime}}\text{.}
\label{pd}%
\end{equation}
We stress that $J$ is the unique positive operator that satisfies Eq.
(\ref{pd}). As a matter of fact, multiplying $A_{k}^{\prime}=U_{k}J$ on the
left by the adjoint equation $A_{k}^{\prime\dagger}=JU_{k}^{\dagger}$ gives,%
\begin{equation}
A_{k}^{\prime\dagger}A_{k}^{\prime}=JU_{k}^{\dagger}U_{k}J=J^{2}\Rightarrow
J=\sqrt{A_{k}^{\prime\dagger}A_{k}^{\prime}}\text{.}%
\end{equation}
Furthermore, if $A_{k}^{\prime}$ is invertible (that is, $\det A_{k}^{\prime
}\neq0$), $U_{k}$ is unique and\textbf{ }reads,%
\begin{equation}
U_{k}\overset{\text{def}}{=}A_{k}^{\prime}J^{-1}=A_{k}^{\prime}\left(
\sqrt{A_{k}^{\prime\dagger}A_{k}^{\prime}}\right)  ^{-1}\text{.} \label{ai}%
\end{equation}
How do we choose the unitary $U_{k}$ when $A_{k}^{\prime}$ is not invertible?
The operator $J$ is a positive operator and belongs to a special subclass of
Hermitian operators such that for any vector $\left\vert v\right\rangle \in
V$, $\left\langle v\left\vert J\right\vert v\right\rangle $ is a \emph{real}
and non-negative number. Therefore, $J$ has a spectral decomposition%
\begin{equation}
J\overset{\text{def}}{=}\sqrt{A_{k}^{\prime\dagger}A_{k}^{\prime}}=%
{\displaystyle\sum\limits_{l}}
\lambda_{l}\left\vert l\right\rangle \left\langle l\right\vert \text{,}%
\end{equation}
where $\lambda_{l}\geq0$ and $\left\{  \left\vert l\right\rangle \right\}  $
denotes an orthonormal basis for the vector space $V$. Define the vectors
$\left\vert \psi_{l}\right\rangle \overset{\text{def}}{=}A_{k}^{\prime
}\left\vert l\right\rangle $ and notice that,%
\begin{equation}
\left\langle \psi_{l}\left\vert \psi_{l}\right.  \right\rangle =\left\langle
l\left\vert A_{k}^{\prime\dagger}A_{k}^{\prime}\right\vert l\right\rangle
=\left\langle l\left\vert J^{2}\right\vert l\right\rangle =\lambda_{l}%
^{2}\text{.}%
\end{equation}
For the time being, consider only those $l$ for which $\lambda_{l}\neq0$. For
those $l$, consider the vectors $\left\vert e_{l}\right\rangle $ defined as%
\begin{equation}
\left\vert e_{l}\right\rangle \overset{\text{def}}{=}\frac{\left\vert \psi
_{l}\right\rangle }{\lambda_{l}}=\frac{A_{k}^{\prime}\left\vert l\right\rangle
}{\lambda_{l}}\text{,}%
\end{equation}
with $\left\langle e_{l}\left\vert e_{l^{\prime}}\right.  \right\rangle
=\delta_{ll^{\prime}}$. For those $l$ for which $\lambda_{l}=0$, extend the
orthonormal set $\left\{  \left\vert e_{l}\right\rangle \right\}  $ in such a
manner that it forms an orthonormal basis $\left\{  \left\vert E_{l}%
\right\rangle \right\}  $. Then, a suitable choice for the unitary operator
$U_{k}$ such that%
\begin{equation}
A_{k}^{\prime}\left\vert l\right\rangle =U_{k}J\left\vert l\right\rangle
\text{,}%
\end{equation}
with $\left\{  \left\vert l\right\rangle \right\}  $ an orthonormal basis for
$V$ \ reads,%
\begin{equation}
U_{k}\overset{\text{def}}{=}%
{\displaystyle\sum\limits_{l}}
\left\vert E_{l}\right\rangle \left\langle l\right\vert \text{.} \label{ani}%
\end{equation}
In summary, the unitary $U_{k}$ is uniquely determined by Eq. (\ref{ai}) when
$A_{k}^{\prime}$ is invertible or Eq. (\ref{ani}) when $A_{k}^{\prime}$ is not
necessarily invertible. We finally stress that the non-uniqueness of $U_{k}$
when $\det A_{k}^{\prime}=0$ is due to the freedom in choosing the orthonormal
basis $\left\{  \left\vert l\right\rangle \right\}  $ for the vector space $V$.

In the scenario being considered, when Eq. (\ref{vai2}) is satisfied, the
enlarged error operators $A_{m}^{\prime}$ admit polar decompositions,%
\begin{equation}
A_{m}^{\prime}P_{\mathcal{C}}=\sqrt{p_{m}}U_{m}P_{\mathcal{C}}\text{,}
\label{pd1}%
\end{equation}
with $k\in\mathcal{K}_{\text{reversible}}$. From Eqs. (\ref{vai2}) and
(\ref{pd1}), we get%
\begin{equation}
p_{m}\delta_{lm}P_{\mathcal{C}}=P_{\mathcal{C}}A_{l}^{\prime\dagger}%
A_{m}^{\prime}P_{\mathcal{C}}=\sqrt{p_{l}p_{m}}P_{\mathcal{C}}U_{l}^{\dagger
}U_{m}P_{\mathcal{C}}\text{,}%
\end{equation}
that is,%
\begin{equation}
P_{\mathcal{C}}U_{l}^{\dagger}U_{m}P_{\mathcal{C}}=\delta_{lm}P_{\mathcal{C}%
}\text{.} \label{condo}%
\end{equation}
We stress that Eq. (\ref{condo}) is needed for an unambiguous syndrome
detection since, as a consequence of the orthogonality of different
$R_{m}^{\dagger}\overset{\text{def}}{=}U_{m}P_{\mathcal{C}}$, the recovery
operation $\mathcal{R}\overset{\text{def}}{=}\left\{  R_{m}\right\}  $ is
trace preserving. This can be shown as follows.

Let $\mathcal{V}^{i_{L}}$ be the subspace of $\mathcal{H}_{2}^{n}$ spanned by
the corrupted images $\left\{  A_{k}^{\prime}\left\vert i_{L}\right\rangle
\right\}  $ of the codewords $\left\vert i_{L}\right\rangle $. Let $\left\{
\left\vert v_{r}^{i_{L}}\right\rangle \right\}  $ be an orthonormal basis for
$\mathcal{V}^{i_{L}}$. We define such a subspace $\mathcal{V}^{i_{L}}$ for
each of the codewords. Because of the KL-conditions \cite{knill-laflamme},%
\begin{align}
\left\langle i_{L}|A_{k}^{\dagger}A_{k^{\prime}}|i_{L}\right\rangle  &
=\left\langle j_{L}|A_{k}^{\dagger}A_{k^{\prime}}|j_{L}\right\rangle \text{,
}\forall i\text{, }j\nonumber\\
& \nonumber\\
\left\langle i_{L}|A_{k}^{\dagger}A_{k^{\prime}}|j_{L}\right\rangle  &
=0\text{, }\forall\text{ }i\neq j\text{,}%
\end{align}
the subspaces $\mathcal{V}^{i_{L}}$ and $\mathcal{V}^{j_{L}}$ with $i\neq j$
are orthogonal subspaces. If $\mathcal{V}^{i_{L}}\oplus\mathcal{V}^{j_{L}}$ is
a proper subset of $\mathcal{H}_{2}^{n}$ with $\mathcal{V}^{i_{L}}%
\oplus\mathcal{V}^{j_{L}}\neq\mathcal{H}_{2}^{n}$, we denote its orthogonal
complement by $\mathcal{O}$. We then have,%
\begin{equation}
\mathcal{H}_{2}^{n}\overset{\text{def}}{=}\left(  \mathcal{V}^{i_{L}}%
\oplus\mathcal{V}^{j_{L}}\right)  \oplus\left(  \mathcal{V}^{i_{L}}%
\oplus\mathcal{V}^{j_{L}}\right)  ^{\perp}=\left(  \mathcal{V}^{i_{L}}%
\oplus\mathcal{V}^{j_{L}}\right)  \oplus\mathcal{O}\text{,}%
\end{equation}
where,%
\begin{equation}
\mathcal{O}\overset{\text{def}}{=}\left(  \mathcal{V}^{i_{L}}\oplus
\mathcal{V}^{j_{L}}\right)  ^{\perp}\text{.} \label{deo}%
\end{equation}
Let $\left\{  \left\vert o_{k}\right\rangle \right\}  $ be an orthonormal
basis for $\mathcal{O}$. Then, the set of states $\left\{  \left\vert
v_{r}^{i_{L}}\right\rangle \text{, }\left\vert o_{k}\right\rangle \right\}  $
constitutes an orthonormal basis for $\mathcal{H}_{2}^{n}$. We introduce the
quantum recovery operation $\mathcal{R}$ with operation elements%
\begin{equation}
\mathcal{R}\overset{\text{def}}{=}\left\{  R_{1}\text{,..., }R_{r}\text{,...,
}\hat{O}\right\}  \text{,}%
\end{equation}
with,%
\begin{equation}
\mathcal{R}\left(  \rho\right)  \overset{\text{def}}{=}\sum_{k\in
\mathcal{K}_{\text{reversible}}^{\prime}}R_{k}\rho R_{k}^{\dagger}+\hat{O}%
\rho\hat{O}^{\dagger}\text{,}%
\end{equation}
where,%
\begin{equation}
R_{r}\overset{\text{def}}{=}\sum_{i}\left\vert i_{L}\right\rangle \left\langle
v_{r}^{i_{L}}\right\vert \text{,} \label{recoveryb}%
\end{equation}
and $\hat{O}$ (with $\hat{O}=\hat{O}^{\dagger}=\hat{O}^{\dagger}\hat{O}$) is a
projector onto the subspace $\mathcal{O}$ in Eq. (\ref{deo}),%
\begin{equation}
\hat{O}\overset{\text{def}}{=}\sum_{k}\left\vert o_{k}\right\rangle
\left\langle o_{k}\right\vert \text{.}%
\end{equation}
We remark that the recovery operation $\mathcal{R}$ is a trace preserving
quantum operation by construction because,%
\begin{align}
\sum_{r}R_{r}^{\dagger}R_{r}+\hat{O}^{\dagger}\hat{O}  &  =\sum_{r}\left[
\left(  \sum_{i}\left\vert i_{L}\right\rangle \left\langle v_{r}^{i_{L}%
}\right\vert \right)  ^{\dagger}\left(  \sum_{j}\left\vert j_{L}\right\rangle
\left\langle v_{r}^{j_{L}}\right\vert \right)  \right]  +\left(  \sum
_{k}\left\vert o_{k}\right\rangle \left\langle o_{k}\right\vert \right)
^{\dagger}\left(  \sum_{k^{\prime}}\left\vert o_{k^{\prime}}\right\rangle
\left\langle o_{k^{\prime}}\right\vert \right) \nonumber\\
& \nonumber\\
&  =\sum_{r\text{, }i\text{, }j}\left\vert v_{r}^{i_{L}}\right\rangle
\left\langle i_{L}|j_{L}\right\rangle \left\langle v_{r}^{j_{L}}\right\vert
+\sum_{k\text{, }k^{\prime}}\left\vert o_{k}\right\rangle \left\langle
o_{k}|o_{k^{\prime}}\right\rangle \left\langle o_{k^{\prime}}\right\vert
\nonumber\\
& \nonumber\\
&  =\sum_{r\text{, }i\text{, }j}\left\vert v_{r}^{i_{L}}\right\rangle
\left\langle v_{r}^{j_{L}}\right\vert \delta_{ij}+\sum_{k\text{, }k^{\prime}%
}\left\vert o_{k}\right\rangle \left\langle o_{k^{\prime}}\right\vert
\delta_{kk^{\prime}}\nonumber\\
& \nonumber\\
&  =\sum_{r\text{, }i}\left\vert v_{r}^{i_{L}}\right\rangle \left\langle
v_{r}^{i_{L}}\right\vert +\sum_{k}\left\vert o_{k}\right\rangle \left\langle
o_{k}\right\vert \nonumber\\
& \nonumber\\
&  =\mathcal{I}_{2^{n}\times2^{n}}\text{,}%
\end{align}
since $\mathcal{B}_{\mathcal{H}_{2}^{n}}\overset{\text{def}}{=}\left\{
\left\vert v_{r}^{j_{L}}\right\rangle \text{, }\left\vert o_{k}\right\rangle
\right\}  $ is an orthonormal basis for $\mathcal{H}_{2}^{n}$. For more
details, we refer to \cite{knill-laflamme}.

\subsection{Approximate quantum error correction conditions}

In general, approximate quantum error correction becomes useful when the
operator-sum representation of the noise model is defined by errors
parametrized by a certain number of small parameters such as the coupling
strength between the environment and the quantum system. For the sake of
simplicity, suppose the error model is characterized by a single small
parameter $\delta$ and assume the goal is to uncover a quantum code for the
noise model $\Lambda^{\prime}$ with fidelity,%
\begin{equation}
\mathcal{F}\geq1-O\left(  \delta^{\beta+1}\right)  \text{,} \label{preserve}%
\end{equation}
for some $\beta\geq0$. How strong can be the violation of the standard (exact)
KL-conditions in order to preserve the desired fidelity order in Eq.
(\ref{preserve})?\ In other words, how relaxed can the approximate error
correction conditions be so that the inequality in (\ref{preserve}) is
satisfied? The answer to this important question was provided by Leung et
\textit{al}. in \cite{leung}.

It turns out that for both exact and approximate quantum error correction
conditions, it is necessary that%
\begin{equation}
P_{\text{detection}}\overset{\text{def}}{=}\sum_{k\in\mathcal{K}%
_{\text{reversible}}^{\prime}}p_{k}\geq\mathcal{F}\text{,} \label{detection}%
\end{equation}
where $P_{\text{detection}}$ denotes the total error detection probability.
Eq. (\ref{detection}) requires that all the enlarged error operators
$A_{l}^{\prime}$ with maximum detection probability must be included in
$\mathcal{A}_{\text{reversible}}^{\prime}$,%
\begin{equation}
\max_{\left\vert \psi_{in}\right\rangle \in\mathcal{C}}\text{Tr}\left(
\left\vert \psi_{in}\right\rangle \left\langle \psi_{in}\right\vert
A_{l}^{\prime\dagger}A_{l}^{\prime}\right)  \approx O\left(  \delta^{\alpha
}\right)  \text{ with }\alpha\leq\beta\text{.}%
\end{equation}
The important point is that a good overlap between the input and output states
is needed while it is not necessary to recover the exact input state
$\left\vert \psi_{in}\right\rangle \left\langle \psi_{in}\right\vert $, since
we do not require $\mathcal{F}=1$. In terms of the enlarged error operators
restricted to the codespace, this means that such errors need to be only
approximately unitary and mutually orthogonal. These considerations lead to
the relaxed sufficient error correction conditions.

In analogy to (\ref{pd1}), assume that the polar decomposition for
$A_{l}^{\prime}$ is given by,%
\begin{equation}
A_{l}^{\prime}P_{\mathcal{C}}=U_{l}\sqrt{P_{\mathcal{C}}A_{l}^{\prime\dagger
}A_{l}^{\prime}P_{\mathcal{C}}}\text{.} \label{sup}%
\end{equation}
Since $P_{\mathcal{C}}A_{l}^{\prime\dagger}A_{l}^{\prime}P_{\mathcal{C}}$
restricted to the codespace $\mathcal{C}$ have different eigenvalues, the
exact error correction conditions are not fulfilled. Let us say that $p_{l}$
and $\lambda_{l}p_{l}$ are the largest and the smallest eigenvalues,
respectively, where both $p_{l}$ and $\lambda_{l}$ are $c$-numbers.
Furthermore, let us define the so-called residue operator $\pi_{l}$ as
\cite{leung},%
\begin{equation}
\pi_{l}\overset{\text{def}}{=}\sqrt{P_{\mathcal{C}}A_{l}^{\prime\dagger}%
A_{l}^{\prime}P_{\mathcal{C}}}-\sqrt{\lambda_{l}p_{l}}P_{\mathcal{C}}\text{,}
\label{pi}%
\end{equation}
where,%
\begin{equation}
0\leq\left\vert \pi_{l}\right\vert \overset{\text{def}}{=}\left(  \pi
_{l}^{\dagger}\pi_{l}\right)  ^{\frac{1}{2}}\leq\sqrt{p_{l}}-\sqrt{\lambda
_{l}p_{l}}\text{.}%
\end{equation}
Substituting (\ref{pi}) into (\ref{sup}), we get%
\begin{equation}
A_{l}^{\prime}P_{\mathcal{C}}=U_{l}\left(  \sqrt{\lambda_{l}p_{l}}I+\pi
_{l}\right)  P_{\mathcal{C}}\text{.} \label{qs}%
\end{equation}
From Eq. (\ref{qs}) and imposing that $P_{\mathcal{C}}U_{l}^{\dagger}%
U_{m}P_{\mathcal{C}}=\delta_{lm}P_{\mathcal{C}}$, the analog of Eq.
(\ref{vai2}) becomes%
\begin{equation}
P_{\mathcal{C}}A_{l}^{\prime\dagger}A_{m}^{\prime}P_{\mathcal{C}}=\left(
\sqrt{\lambda_{l}p_{l}}I+\pi_{l}^{\dagger}\right)  \left(  \sqrt{\lambda
_{m}p_{m}}I+\pi_{m}\right)  P_{\mathcal{C}}\delta_{lm}\text{,} \label{vai3}%
\end{equation}
where,%
\begin{equation}
p_{l}\left(  1-\lambda_{l}\right)  \leq O\left(  \delta^{\beta+1}\right)
\text{, }\forall l\in\mathcal{K}_{\text{reversible}}^{\prime}\text{.}%
\end{equation}
We stress that when the exact error correction conditions are satisfied,
$\lambda_{l}=1$ and $\pi_{l}=0$ (the null operator). Thus, in that scenario,
Eqs. (\ref{vai2}) and (\ref{vai3}) coincide. Finally, we point out that an
approximate recovery operation $\mathcal{R}\overset{\text{def}}{=}\left\{
R_{1}\text{,..., }R_{r}\text{,..., }\hat{O}\right\}  $ with $R_{k}$ defined in
Eq. (\ref{recoveryb})\textbf{ }and $\hat{O}$ formally defined just as in the
exact case can be employed in this new scenario as well. However, assuming to
consider $R_{m}^{\dagger}\overset{\text{def}}{=}U_{m}P_{\mathcal{C}}$, extra
care in the explicit computation of the unitary operators $U_{m}$ is needed in
view of the fact that the polar decomposition (\ref{pd1}) is replaced by the
one in (\ref{sup}). For an explicit unabridged and correct computation of
recovery operators as originally proposed by Leung et \textit{al}., see
Appendix A. For further theoretical details, we refer to reference
\cite{leung}.

\section{From exact to approximate QEC: two simple noise models}

The objective in this section is to discuss in detail the exact and
approximate error correction conditions for the simplest unital and nonunital
channels, respectively. Specifically, we consider the bit-flip and the
amplitude damping noise models. Error correction is performed by means of the
three-qubit bit-flip repetition code for the unital channel while we employ
the four-qubit Leung et \textit{al}. code for the nonunital noise model. We
acknowledge that the bit-flip noise model is certainly not the prototype of a
truly quantum noise model. However, we believe its consideration is suitable
for our purposes, since we wish to essentially stress the similarities and
differences between exact and approximate error correction schemes avoiding
unnecessary complications. The exact error correction analysis for more
realistic and truly quantum error models along the lines presented here could
be found in previous works of one of the Authors \cite{cafaro1, cafaro2,
cafaro3}.

\subsection{The simplest unital channel with Pauli errors}

We consider a bit-flip noisy quantum channel and QEC is performed via the
three-qubit bit-flip repetition code \cite{nielsenbook}. We remark that the
bit-flip and the phase-flip (or, dephasing) channels are unitarily equivalent.
This means that there exists a unitary operator\textbf{ }$U$\textbf{\ }such
that the action of one channel is the same as the other, provided the first
channel is proceed by\textbf{ }$U$\textbf{\ }and followed by\textbf{
}$U^{\dagger}$\textbf{. }In the case being considered, it follows that%
\[
\Lambda_{\text{bit}}\left(  \rho\right)  \overset{\text{def}}{=}\left(
H\circ\Lambda_{\text{phase}}\circ H^{\dagger}\right)  \left(  \rho\right)
\text{,}%
\]
where $H$\textbf{\ }denotes the Hadamard single-qubit gate. Error correction
of dephasing errors by means of the three-qubit phase flip repetition code
works very much like the error correction of bit-flip errors via the
three-qubit bit flip repetition code. That said, we admit that a pure
dephasing channel, with no other sources of noise at all, is physically
improbable. However, in many physical systems, dephasing is indeed the
dominant error source \cite{ben}.

The performance of the error correcting code is quantified by means of the
entanglement fidelity as function of the error probability. The bit flip noisy
channel $\Lambda_{\text{bit}}^{\left(  1\right)  }$ (single use of the
channel) is defined as follows,%
\begin{equation}
\Lambda_{\text{bit}}^{\left(  1\right)  }\left(  \rho\right)  \overset
{\text{def}}{=}\left(  1-p\right)  \rho+pX\rho X^{\dagger}\text{,}%
\end{equation}
where the matrix representation in the $1$-qubit computational basis
$\mathcal{B}_{\text{computational}}=\left\{  \left\vert 0\right\rangle \text{,
}\left\vert 1\right\rangle \right\}  $ of the $X$-Pauli operator is given by,%
\begin{equation}
\left[  X\right]  _{\mathcal{B}_{\text{computational}}}\overset{\text{def}}%
{=}\left(
\begin{array}
[c]{cc}%
\left\langle 0|X|0\right\rangle  & \left\langle 0|X|1\right\rangle \\
\left\langle 1|X|0\right\rangle  & \left\langle 1|X|1\right\rangle
\end{array}
\right)  =\left(
\begin{array}
[c]{cc}%
0 & 1\\
1 & 0
\end{array}
\right)  \text{.}%
\end{equation}
Observe that the bit-flip channel is a unital channel since $\Lambda
_{\text{bit}}^{\left(  1\right)  }\left(  I\right)  =I$. Consider the
three-qubit bit flip encoding defined as,%
\begin{equation}
\left\vert 0\right\rangle \rightarrow\left\vert 0_{L}\right\rangle
\overset{\text{def}}{=}\left\vert 000\right\rangle \text{, }\left\vert
1\right\rangle \rightarrow\left\vert 1_{L}\right\rangle \overset{\text{def}%
}{=}\left\vert 111\right\rangle \text{.}%
\end{equation}
The action of three uses of the bit flip channel $\Lambda_{\text{bit}%
}^{\left(  3\right)  }\left(  \rho\right)  $ on $3$-qubits quantum states
reads,%
\begin{equation}
\Lambda_{\text{bit}}^{\left(  3\right)  }\left(  \rho\right)  \overset
{\text{def}}{=}\sum_{i_{1}\text{, }i_{2}\text{, }i_{3}=0}^{1}p_{i_{3}}%
p_{i_{2}}p_{i_{1}}\left(  A_{i_{3}}\otimes A_{i_{2}}\otimes A_{i_{1}}\right)
\rho\left(  A_{i_{3}}\otimes A_{i_{2}}\otimes A_{i_{1}}\right)  ^{\dagger
}\text{,}%
\end{equation}
where $A_{0}\overset{\text{def}}{=}I$, $A_{1}\overset{\text{def}}{=}X$ are
Pauli operators. Furthermore,%
\begin{equation}
p_{0}\overset{\text{def}}{=}1-p\text{,}\;p_{1}\overset{\text{def}}{=}p\text{,}
\label{aaa1}%
\end{equation}
with,%
\begin{equation}
\sum_{i_{1}\text{, }i_{2}\text{, }i_{3}=0}^{1}p_{i_{3}}p_{i_{2}}p_{i_{1}%
}=p_{0}^{3}+3p_{1}p_{0}^{2}+3p_{1}^{2}p_{0}+p_{1}^{3}=1\text{.}%
\end{equation}
To simplify our notation, we may assume that $A_{i_{n}}\otimes$...$\otimes
A_{i_{1}}\equiv$ $A_{i_{n}}$...$A_{i_{1}}$. The channel $\Lambda_{\text{bit}%
}^{(3)}(\rho)$ can be written as,%
\begin{equation}
\text{ }\Lambda_{\text{bit}}^{(3)}(\rho)\overset{\text{def}}{=}%
{\displaystyle\sum\limits_{k=0}^{7}}
A_{k}^{\prime}\rho A_{k}^{\prime\dagger}\text{ and, }%
{\displaystyle\sum\limits_{k=0}^{7}}
A_{k}^{\prime\dagger}A_{k}^{\prime}=I_{8\times8}\text{,} \label{nota}%
\end{equation}
where we denote with $\mathcal{A}$ the superoperator defined in terms of the
enlarged error operators $\left\{  A_{0}^{\prime}\text{,.., }A_{7}^{\prime
}\right\}  $. In an explicit way, the error operators $\left\{  A_{0}^{\prime
}\text{,.., }A_{7}^{\prime}\right\}  $ read,%
\begin{align}
&  A_{0}^{\prime}\overset{\text{def}}{=}\sqrt{p_{0}^{3}}I^{1}\otimes
I^{2}\otimes I^{3}\text{, }A_{1}^{\prime}\overset{\text{def}}{=}\sqrt
{p_{1}p_{0}^{2}}X^{1}\otimes I^{2}\otimes I^{3}\text{, }A_{2}^{\prime}%
\overset{\text{def}}{=}\sqrt{p_{1}p_{0}^{2}}I^{1}\otimes X^{2}\otimes
I^{3}\text{, }\nonumber\\
& \nonumber\\
&  A_{3}^{\prime}\overset{\text{def}}{=}\sqrt{p_{1}p_{0}^{2}}I^{1}\otimes
I^{2}\otimes X^{3}\text{, }A_{4}^{\prime}\overset{\text{def}}{=}\sqrt
{p_{1}^{2}p_{0}}X^{1}\otimes X^{2}\otimes I^{3}\text{, }A_{5}^{\prime}%
\overset{\text{def}}{=}\sqrt{p_{1}^{2}p_{0}}X^{1}\otimes I^{2}\otimes
X^{3}\text{,}\nonumber\\
& \nonumber\\
&  \text{ }A_{6}^{\prime}\overset{\text{def}}{=}\sqrt{p_{1}^{2}p_{0}}%
I^{1}\otimes X^{2}\otimes X^{3}\text{, }A_{7}^{\prime}\overset{\text{def}}%
{=}\sqrt{p_{1}^{3}}X^{1}\otimes X^{2}\otimes X^{3}\text{.} \label{fuck1}%
\end{align}
The set of error operators satisfying the detectability condition,
$P_{\mathcal{C}}A_{k}^{\prime}P_{\mathcal{C}}=\lambda_{A_{k}^{\prime}%
}P_{\mathcal{C}}$, where $P_{\mathcal{C}}\overset{\text{def}}{=}\left\vert
0_{L}\right\rangle \left\langle 0_{L}\right\vert +$ $\left\vert 1_{L}%
\right\rangle \left\langle 1_{L}\right\vert $ is the projector operator on the
code subspace $\mathcal{C}\overset{\text{def}}{=}$Span$\left\{  \left\vert
0_{L}\right\rangle \text{, }\left\vert 1_{L}\right\rangle \right\}  $ is given
by,%
\begin{equation}
\mathcal{A}_{\text{detectable}}\overset{\text{def}}{=}\left\{  A_{0}^{\prime
}\text{, }A_{1}^{\prime}\text{, }A_{2}^{\prime}\text{, }A_{3}^{\prime}\text{,
}A_{4}^{\prime}\text{, }A_{5}^{\prime}\text{, }A_{6}^{\prime}\right\}
\subseteq\mathcal{A}\text{.}%
\end{equation}
The only non detectable error is $A_{7}^{\prime}$. Furthermore, since all the
detectable errors are invertible, the set of correctable errors is such that
$\mathcal{A}_{\text{correctable}}^{\dagger}\mathcal{A}_{\text{correctable}}$
is detectable \cite{zurek}. It follows that,%
\begin{equation}
\mathcal{A}_{\text{correctable}}\overset{\text{def}}{=}\left\{  A_{0}^{\prime
}\text{, }A_{1}^{\prime}\text{, }A_{2}^{\prime}\text{, }A_{3}^{\prime
}\right\}  \subseteq\mathcal{A}_{\text{detectable}}\subseteq\mathcal{A}%
\text{.}%
\end{equation}
To be more explicit, the set of enlarged error operators $\left\{
A_{k}^{\prime}\right\}  $ with $k\in\left\{  0\text{,..., }\bar{k}\right\}  $
is correctable provided that,%
\begin{equation}
P_{\mathcal{C}}A_{l}^{\prime\dagger}A_{m}^{\prime}P_{\mathcal{C}}\propto
P_{\mathcal{C}}\text{,} \label{peq}%
\end{equation}
for any pair of $\left(  l\text{, }m\right)  $ with $l$, $m\in\left\{
0\text{,..., }\bar{k}\right\}  $. Eq. (\ref{peq}) is satisfied if and only if,%
\begin{equation}
\left\langle 0_{L}\left\vert A_{l}^{\prime\dagger}A_{m}^{\prime}\right\vert
0_{L}\right\rangle =\left\langle 1_{L}\left\vert A_{l}^{\prime\dagger}%
A_{m}^{\prime}\right\vert 1_{L}\right\rangle \text{ and, }\left\langle
0_{L}\left\vert A_{l}^{\prime\dagger}A_{m}^{\prime}\right\vert 1_{L}%
\right\rangle =\left\langle 1_{L}\left\vert A_{l}^{\prime\dagger}A_{m}%
^{\prime}\right\vert 0_{L}\right\rangle =0\text{,}%
\end{equation}
for any pair of $\left(  l\text{, }m\right)  $ with $l$, $m$ $\in\left\{
0\text{,..., }\bar{k}\right\}  $. The enlarged error operators $\left\{
A_{l}^{\prime}\right\}  $ in (\ref{fuck1}) can be rewritten as,%
\begin{align}
&  A_{0}^{\prime}\overset{\text{def}}{=}\sqrt{\left(  1-p\right)  ^{3}%
}I\text{, }A_{1}^{\prime}\overset{\text{def}}{=}\sqrt{p\left(  1-p\right)
^{2}}X^{1}\text{, }A_{2}^{\prime}\overset{\text{def}}{=}\sqrt{p\left(
1-p\right)  ^{2}}X^{2}\text{,}\nonumber\\
& \nonumber\\
&  A_{3}^{\prime}\overset{\text{def}}{=}\sqrt{p\left(  1-p\right)  ^{2}}%
X^{3}\text{, }A_{4}^{\prime}\overset{\text{def}}{=}\sqrt{p^{2}\left(
1-p\right)  }X^{1}X^{2}\text{, }A_{5}^{\prime}\overset{\text{def}}{=}%
\sqrt{p^{2}\left(  1-p\right)  }X^{1}X^{3}\text{, }\nonumber\\
& \nonumber\\
&  A_{6}^{\prime}\overset{\text{def}}{=}\sqrt{p^{2}\left(  1-p\right)  }%
X^{2}X^{3}\text{, }A_{7}^{\prime}\overset{\text{def}}{=}\sqrt{p^{3}}X^{1}%
X^{2}X^{3}\text{.} \label{enlarged}%
\end{align}
The action of the correctable error operators $\mathcal{A}_{\text{correctable}%
}$ on the codewords $\left\vert 0_{L}\right\rangle $ and $\left\vert
1_{L}\right\rangle $ is given by,%
\begin{align}
\left\vert 0_{L}\right\rangle  &  \rightarrow A_{0}^{\prime}\left\vert
0_{L}\right\rangle =\sqrt{p_{0}^{3}}\left\vert 000\right\rangle \text{, }%
A_{1}^{\prime}\left\vert 0_{L}\right\rangle =\sqrt{p_{1}p_{0}^{2}}\left\vert
100\right\rangle \text{, }A_{2}^{\prime}\left\vert 0_{L}\right\rangle
=\sqrt{p_{1}p_{0}^{2}}\left\vert 010\right\rangle \text{, }A_{3}^{\prime
}\left\vert 0_{L}\right\rangle =\sqrt{p_{1}p_{0}^{2}}\left\vert
001\right\rangle \text{ }\nonumber\\
& \nonumber\\
\left\vert 1_{L}\right\rangle  &  \rightarrow A_{0}^{\prime}\left\vert
1_{L}\right\rangle =\sqrt{p_{0}^{3}}\left\vert 111\right\rangle \text{, }%
A_{1}^{\prime}\left\vert 1_{L}\right\rangle =\sqrt{p_{1}p_{0}^{2}}\left\vert
011\right\rangle \text{, }A_{2}^{\prime}\left\vert 1_{L}\right\rangle
=\sqrt{p_{1}p_{0}^{2}}\left\vert 101\right\rangle \text{, }A_{3}^{\prime
}\left\vert 1_{L}\right\rangle =\sqrt{p_{1}p_{0}^{2}}\left\vert
110\right\rangle \text{.} \label{ea}%
\end{align}
The two \ four-dimensional orthogonal subspaces $\mathcal{V}^{0_{L}}$ and
$\mathcal{V}^{1_{L}}$ of $\mathcal{H}_{2}^{3}$ generated by the action of
$\mathcal{A}_{\text{correctable}}$ on $\left\vert 0_{L}\right\rangle $ and
$\left\vert 1_{L}\right\rangle $ are defined as,%
\begin{equation}
\mathcal{V}^{0_{L}}\overset{\text{def}}{=}\text{Span}\left\{  \left\vert
v_{1}^{0_{L}}\right\rangle =\left\vert 000\right\rangle \text{,}\left\vert
v_{2}^{0_{L}}\right\rangle =\left\vert 100\right\rangle \text{, }\left\vert
v_{3}^{0_{L}}\right\rangle =\left\vert 010\right\rangle \text{, }\left\vert
v_{4}^{0_{L}}\right\rangle =\left\vert 001\right\rangle \text{ }\right\}
\text{,} \label{span1}%
\end{equation}
and,%
\begin{equation}
\mathcal{V}^{1_{L}}\overset{\text{def}}{=}\text{Span}\left\{  \left\vert
v_{1}^{1_{L}}\right\rangle =\left\vert 111\right\rangle \text{,}\left\vert
v_{2}^{1_{L}}\right\rangle =\left\vert 011\right\rangle \text{, }\left\vert
v_{3}^{1_{L}}\right\rangle =\left\vert 101\right\rangle \text{, }\left\vert
v_{4}^{1_{L}}\right\rangle =\left\vert 110\right\rangle \right\}  \text{,}
\label{span2}%
\end{equation}
respectively. Notice that $\mathcal{V}^{0_{L}}\oplus\mathcal{V}^{1_{L}%
}=\mathcal{H}_{2}^{3}$. The recovery superoperator $\mathcal{R}\leftrightarrow
\left\{  R_{l}\right\}  $ with $l=1$,.., $4$ is defined as
\cite{knill-laflamme},%
\begin{equation}
R_{l}\overset{\text{def}}{=}V_{l}\sum_{i=0}^{1}\left\vert v_{l}^{i_{L}%
}\right\rangle \left\langle v_{l}^{i_{L}}\right\vert \text{,} \label{recovery}%
\end{equation}
where the unitary operator $V_{l}$ is such that $V_{l}\left\vert v_{l}^{i_{L}%
}\right\rangle \overset{\text{def}}{=}\left\vert i_{L}\right\rangle $ for
$i\in\left\{  0\text{, }1\right\}  $. Substituting (\ref{span1}) and
(\ref{span2}) into (\ref{recovery}), it follows that the four recovery
operators $\left\{  R_{0}\text{, }R_{1}\text{, }R_{2}\text{, }R_{3}\right\}  $
are given by,%
\begin{align}
R_{0}\overset{\text{def}}{=}\frac{1}{\sqrt{\left(  1-p\right)  ^{3}}%
}P_{\mathcal{C}}A_{0}^{\prime\dagger}  &  =\left\vert 000\right\rangle
\left\langle 000\right\vert +\left\vert 111\right\rangle \left\langle
111\right\vert \text{, }R_{1}\overset{\text{def}}{=}\frac{1}{\sqrt{p\left(
1-p\right)  ^{2}}}P_{\mathcal{C}}A_{1}^{\prime\dagger}=\left\vert
000\right\rangle \left\langle 100\right\vert +\left\vert 111\right\rangle
\left\langle 011\right\vert \text{,}\nonumber\\
& \nonumber\\
R_{2}\overset{\text{def}}{=}\frac{1}{\sqrt{p\left(  1-p\right)  ^{2}}%
}P_{\mathcal{C}}A_{2}^{\prime\dagger}  &  =\left\vert 000\right\rangle
\left\langle 010\right\vert +\left\vert 111\right\rangle \left\langle
101\right\vert \text{, }R_{3}\overset{\text{def}}{=}\frac{1}{\sqrt{p^{3}}%
}P_{\mathcal{C}}A_{3}^{\prime\dagger}=\left\vert 000\right\rangle \left\langle
001\right\vert +\left\vert 111\right\rangle \left\langle 110\right\vert
\text{,}%
\end{align}
with,%
\begin{equation}%
{\displaystyle\sum\limits_{l=0}^{3}}
R_{l}^{\dagger}R_{l}=I_{2^{3}\times2^{3}}%
\end{equation}
We observe that the four recovery operators $R_{j}$ associated with the four
correctable errors $A_{j}^{\prime}$ with $j\in\left\{  0\text{, }1\text{,
}2\text{, }3\right\}  $ are formally defined as,%
\begin{equation}
R_{j}\overset{\text{def}}{=}\frac{\left\vert 0_{L}\right\rangle \left\langle
0_{L}\right\vert A_{j}^{\prime\dagger}}{\sqrt{\left\langle 0_{L}\left\vert
A_{j}^{\prime\dagger}A_{j}^{\prime}\right\vert 0_{L}\right\rangle }}%
+\frac{\left\vert 1_{L}\right\rangle \left\langle 1_{L}\right\vert
A_{j}^{\prime\dagger}}{\sqrt{\left\langle 1_{L}\left\vert A_{j}^{\prime
\dagger}A_{j}^{\prime}\right\vert 1_{L}\right\rangle }}\text{.} \label{rj}%
\end{equation}
When considering exact quantum error correction, $R_{j}\propto P_{\mathcal{C}%
}A_{j}^{\prime\dagger}$ where the coefficient of proportionality must be
determined in such a manner that its product with $A_{j}^{\prime\dagger}$
leads to a unitary operator. This coefficient equals the square root of
$\left\langle 0_{L}\left\vert A_{j}^{\prime\dagger}A_{j}^{\prime}\right\vert
0_{L}\right\rangle $ (or, $\left\langle 1_{L}\left\vert A_{j}^{\prime\dagger
}A_{j}^{\prime}\right\vert 1_{L}\right\rangle $). We emphasize three features
of exact-QEC, two of which concern the standard QEC recovery
superoperator\textbf{ }$\mathcal{R}\overset{\text{def}}{=}$\textbf{ }$\left\{
R_{0}\text{, }R_{1}\text{, }R_{2}\text{, }R_{3}\right\}  $:

\begin{itemize}
\item The two eigenvalues $\lambda_{\max}$ and $\lambda_{\min}$ of the
$\left(  2\times2\right)  $-matrix associated with the operators
$P_{\mathcal{C}}A_{l}^{\prime\dagger}A_{m}^{\prime}P_{\mathcal{C}}$ (with
$A_{k}^{\prime}$ correctable errors) on the codespace $\mathcal{C}$ coincide.
For example, for $P_{\mathcal{C}}A_{0}^{\prime\dagger}A_{0}^{\prime
}P_{\mathcal{C}}$ we have $\lambda_{\max}=\lambda_{\min}=\left(  1-p\right)
^{3}$;

\item The projector on the codespace $P_{\mathcal{C}}$ belongs to
$\mathcal{R}$, the standard QEC recovery;

\item All the four recovery operators in $\mathcal{R}$ are $p$-independent,
where $p$ denotes the error probability and is the single parameter that
characterizes the noise model being considered.
\end{itemize}

For the simple bit-flip noise model with error correction performed by means
of the three-qubit bit-flip code, the entanglement fidelity reads \cite{benji,
mike},%
\begin{equation}
\mathcal{F}_{\left[  \left[  3,1,1\right]  \right]  }\left(  p\right)
\overset{\text{def}}{=}\frac{1}{\left(  \dim_{%
\mathbb{C}
}\mathcal{C}\right)  ^{2}}\sum_{l=0}^{7}%
{\displaystyle\sum\limits_{k=0}^{3}}
\left\vert \text{Tr}\left(  R_{k}A_{l}^{\prime}\right)  _{\mathcal{C}%
}\right\vert ^{2}=\frac{1}{4}\sum_{l=0}^{7}%
{\displaystyle\sum\limits_{k=0}^{3}}
\left\vert \left\langle 0_{L}\left\vert R_{k}A_{l}^{\prime}\right\vert
0_{L}\right\rangle +\left\langle 1_{L}\left\vert R_{k}A_{l}^{\prime
}\right\vert 1_{L}\right\rangle \right\vert ^{2}\text{.} \label{ef}%
\end{equation}
Substituting (\ref{rj}) into the RHS of (\ref{ef}), we obtain%
\begin{equation}
\mathcal{F}_{\left[  \left[  3,1,1\right]  \right]  }\left(  p\right)
=\frac{1}{4}\sum_{l=0}^{7}%
{\displaystyle\sum\limits_{k=0}^{3}}
\left\vert \frac{\left\langle 0_{L}\left\vert A_{k}^{\prime\dagger}%
A_{l}^{\prime}\right\vert 0_{L}\right\rangle }{\sqrt{\left\langle
0_{L}\left\vert A_{k}^{\prime\dagger}A_{k}^{\prime}\right\vert 0_{L}%
\right\rangle }}+\frac{\left\langle 1_{L}\left\vert A_{k}^{\prime\dagger}%
A_{l}^{\prime}\right\vert 1_{L}\right\rangle }{\sqrt{\left\langle
1_{L}\left\vert A_{k}^{\prime\dagger}A_{k}^{\prime}\right\vert 1_{L}%
\right\rangle }}\right\vert ^{2}\text{.}%
\end{equation}
We observe that $\mathcal{F}_{\left[  \left[  3,1,1\right]  \right]  }\left(
p\right)  $ is, in principle, the sum of $4\times8=32$-terms that arise by
considering all the possible pairs $\left(  k\text{, }l\right)  $ with
$k\in\left\{  0\text{, }1\text{, }2\text{, }3\right\}  $ and $l\in\left\{
0\text{, }1\text{, }2\text{, }3\text{, }4\text{, }5\text{, }6\text{, }7\text{,
}8\right\}  $. However, it turns out that only $4$-terms are nonvanishing and
contribute to the computation of the entanglement fidelity. They are $\left\{
\left(  k\text{, }l\right)  \right\}  =\left\{  \left(  0\text{, }0\right)
\text{, }\left(  1\text{, }1\right)  \text{, }\left(  2\text{, }2\right)
\text{, }\left(  3\text{, }3\right)  \right\}  $. Thus, only the four
correctable errors $\left\{  A_{0}^{\prime}\text{, }A_{1}^{\prime}\text{,
}A_{2}^{\prime}\text{, }A_{3}^{\prime}\right\}  $ are recoverable. Indeed,
they are fully recovered and,%
\begin{equation}
\left(  R_{j}A_{j}^{\prime\dagger}\right)  _{\mathcal{C}}=\left\langle
0_{L}\left\vert A_{j}^{\prime\dagger}A_{j}^{\prime}\right\vert 0_{L}%
\right\rangle I=\left\langle 1_{L}\left\vert A_{j}^{\prime\dagger}%
A_{j}^{\prime}\right\vert 1_{L}\right\rangle I\propto I\text{.}%
\end{equation}
In the exact-QEC scenario, we stress:

\begin{itemize}
\item Only the correctable errors are recoverable. Indeed, they are fully
recoverable. No off-diagonal contribution arises.
\end{itemize}

In summary, $\mathcal{F}_{\left[  \left[  3,1,1\right]  \right]  }\left(
p\right)  $ reads%
\begin{align}
\mathcal{F}_{\left[  \left[  3,1,1\right]  \right]  }\left(  p\right)   &
=\frac{1}{4}\left[
\begin{array}
[c]{c}%
\left\vert \frac{\left\langle 0_{L}\left\vert A_{0}^{\prime\dagger}%
A_{0}^{\prime}\right\vert 0_{L}\right\rangle }{\sqrt{\left\langle
0_{L}\left\vert A_{0}^{\prime\dagger}A_{0}^{\prime}\right\vert 0_{L}%
\right\rangle }}+\frac{\left\langle 1_{L}\left\vert A_{0}^{\prime\dagger}%
A_{0}^{\prime}\right\vert 1_{L}\right\rangle }{\sqrt{\left\langle
1_{L}\left\vert A_{0}^{\prime\dagger}A_{0}^{\prime}\right\vert 1_{L}%
\right\rangle }}\right\vert ^{2}+\left\vert \frac{\left\langle 0_{L}\left\vert
A_{1}^{\prime\dagger}A_{1}^{\prime}\right\vert 0_{L}\right\rangle }%
{\sqrt{\left\langle 0_{L}\left\vert A_{1}^{\prime\dagger}A_{1}^{\prime
}\right\vert 0_{L}\right\rangle }}+\frac{\left\langle 1_{L}\left\vert
A_{1}^{\prime\dagger}A_{1}^{\prime}\right\vert 1_{L}\right\rangle }%
{\sqrt{\left\langle 1_{L}\left\vert A_{1}^{\prime\dagger}A_{1}^{\prime
}\right\vert 1_{L}\right\rangle }}\right\vert ^{2}+\\
\\
\left\vert \frac{\left\langle 0_{L}\left\vert A_{2}^{\prime\dagger}%
A_{2}^{\prime}\right\vert 0_{L}\right\rangle }{\sqrt{\left\langle
0_{L}\left\vert A_{2}^{\prime\dagger}A_{2}^{\prime}\right\vert 0_{L}%
\right\rangle }}+\frac{\left\langle 1_{L}\left\vert A_{2}^{\prime\dagger}%
A_{2}^{\prime}\right\vert 1_{L}\right\rangle }{\sqrt{\left\langle
1_{L}\left\vert A_{2}^{\prime\dagger}A_{2}^{\prime}\right\vert 1_{L}%
\right\rangle }}\right\vert ^{2}+\left\vert \frac{\left\langle 0_{L}\left\vert
A_{3}^{\prime\dagger}A_{3}^{\prime}\right\vert 0_{L}\right\rangle }%
{\sqrt{\left\langle 0_{L}\left\vert A_{3}^{\prime\dagger}A_{3}^{\prime
}\right\vert 0_{L}\right\rangle }}+\frac{\left\langle 1_{L}\left\vert
A_{4}^{\prime\dagger}A_{4}^{\prime}\right\vert 1_{L}\right\rangle }%
{\sqrt{\left\langle 1_{L}\left\vert A_{4}^{\prime\dagger}A_{4}^{\prime
}\right\vert 1_{L}\right\rangle }}\right\vert ^{2}%
\end{array}
\right] \nonumber\\
& \nonumber\\
&  =\frac{1}{4}\left[  4\left\langle 0_{L}\left\vert A_{0}^{\prime\dagger
}A_{0}^{\prime}\right\vert 0_{L}\right\rangle +4\left\langle 0_{L}\left\vert
A_{1}^{\prime\dagger}A_{1}^{\prime}\right\vert 0_{L}\right\rangle
+4\left\langle 0_{L}\left\vert A_{2}^{\prime\dagger}A_{2}^{\prime}\right\vert
0_{L}\right\rangle +4\left\langle 0_{L}\left\vert A_{3}^{\prime\dagger}%
A_{3}^{\prime}\right\vert 0_{L}\right\rangle \right]  \text{,}%
\end{align}
that is,%
\begin{equation}
\mathcal{F}_{\left[  \left[  3,1,1\right]  \right]  }\left(  p\right)
=\left\langle 0_{L}\left\vert A_{0}^{\prime\dagger}A_{0}^{\prime}\right\vert
0_{L}\right\rangle +\left\langle 0_{L}\left\vert A_{1}^{\prime\dagger}%
A_{1}^{\prime}\right\vert 0_{L}\right\rangle +\left\langle 0_{L}\left\vert
A_{2}^{\prime\dagger}A_{2}^{\prime}\right\vert 0_{L}\right\rangle
+\left\langle 0_{L}\left\vert A_{3}^{\prime\dagger}A_{3}^{\prime}\right\vert
0_{L}\right\rangle \text{.} \label{chist}%
\end{equation}
Substituting (\ref{enlarged}) into (\ref{chist}), we get%
\begin{equation}
\mathcal{F}_{\left[  \left[  3,1,1\right]  \right]  }\left(  p\right)
=1-3p^{2}+2p^{3}\text{.}%
\end{equation}
When the error probability $p$ increases beyond a certain threshold $\bar{p}$,
quantum error correction does more harm than good. To uncover this point
$\bar{p}$, we have to \ check two conditions. First, we compare $\mathcal{F}%
_{\left[  \left[  3,1,1\right]  \right]  }\left(  p\right)  $ with the
fidelity without coding and error correction (the so-called single-qubit
baseline performance),%
\begin{equation}
\mathcal{F}_{\text{no-QEC}}\left(  p\right)  \overset{\text{def}}{=}\frac
{1}{4}%
{\displaystyle\sum\limits_{k=0}^{1}}
\left\vert \text{Tr}A_{k}\right\vert ^{2}=1-2p+p^{2}\text{.}%
\end{equation}
For this error correction scheme to be useful, it must be%
\begin{equation}
\mathcal{F}_{\text{no-QEC}}\left(  p\right)  \leq\mathcal{F}_{\left[  \left[
3,1,1\right]  \right]  }\left(  p\right)  \text{.}%
\end{equation}
In the case being considered, this inequality holds true for any $0\leq
p\leq1$. Second, the error correction scheme is effective provided that the
failure probability $P_{\text{failure}}\left(  p\right)  \overset{\text{def}%
}{=}1-\mathcal{F}_{\left[  \left[  3,1,1\right]  \right]  }\left(  p\right)  $
is smaller than the error probability $p$,%
\begin{equation}
P_{\text{failure}}\left(  p\right)  \leq p\text{.}%
\end{equation}
This second inequality holds true if and only if $0\leq p\leq\frac{1}{2}$.

\subsection{The simplest nonunital channel with non-Pauli errors}

An approximate QEC framework turns out to be of great use when combatting
non-Pauli errors. Within such framework, we allow for a negligible but
non-vanishing error in the recovery so that errors need not be exactly
orthogonal to be unambiguously detected and perfectly recovered. Indeed, we
allow for slight non-orthogonalities between approximately correctable error
operators. This way, correctable errors have to satisfy the KL-conditions only
approximately. As a consequence, in such a scenario, the composite operation
$\mathcal{R}\Lambda\mathcal{E}$ is necessarily only approximately close to the
identity on the codespace. Observe that in such approximate QEC framework, for
a given noise model, more codes satisfying the approximate KL-conditions can
be constructed. Furthermore, it is not unusual to uncover codes of shorter
block lengths which, although of less general applicability, may indeed be
more efficient for the specific error model considered. For instance, when
considering amplitude damping errors in the standard $A_{0}$, $A_{1}$
non-Pauli error basis on a $n$-qubits state, to the first order in $\gamma$,
$\left(  n+1\right)  $-errors may occur. Thus, in order to be correctable by
this nondegenerate non-Pauli basis code, such errors must map the codeword
space to orthogonal spaces if the syndrome is to be detected unambiguously.
Thus, it must be $2^{n}\geq2\left(  n+1\right)  $, that is $n\geq3$. Instead,
considering the error correction of the same decoherence model by means of
nondegenerate standard Pauli basis codes, it must be $2^{n}\geq2\left(
2n+1\right)  $, that is $n\geq5$. The former scenario arises when considering
the amplitude damping channel and using the Leung et \textit{al}. $\left[
\left[  4\text{, }1\right]  \right]  $ code.

In the case of amplitude damping, we model the environment as starting in the
$\left\vert 0\right\rangle $ state as it were at zero temperature. The AD
quantum noisy channel is defined as \cite{nielsenbook},%
\begin{equation}
\Lambda_{\text{AD}}\left(  \rho\right)  \overset{\text{def}}{=}\sum_{k=0}%
^{1}A_{k}\rho A_{k}^{\dagger}\text{,}%
\end{equation}
where the Kraus error operators $A_{k}$ read,
\begin{equation}
A_{0}\overset{\text{def}}{=}\frac{1}{2}\left[  \left(  1+\sqrt{1-\gamma
}\right)  I+\left(  1-\sqrt{1-\gamma}\right)  \sigma_{z}\right]  \text{,
}A_{1}\overset{\text{def}}{=}\frac{\sqrt{\gamma}}{2}\left(  \sigma_{x}%
+i\sigma_{y}\right)  \text{.}%
\end{equation}
Observe that the AD channel is nonunital since $\Lambda_{\text{AD}}\left(
I\right)  \neq I$. The $\left(  2\times2\right)  $-matrix representation of
the $A_{k}$ operators is given by,%
\begin{equation}
A_{0}=\left(
\begin{array}
[c]{cc}%
1 & 0\\
0 & \sqrt{1-\gamma}%
\end{array}
\right)  \text{ and, }A_{1}=\left(
\begin{array}
[c]{cc}%
0 & \sqrt{\gamma}\\
0 & 0
\end{array}
\right)  \text{.}%
\end{equation}
The action of the $A_{k}$ with $k\in\left\{  0\text{, }1\right\}  $ operators
on the computational basis vectors $\left\vert 0\right\rangle $ and
$\left\vert 1\right\rangle $ reads,%
\begin{equation}
A_{0}\left\vert 0\right\rangle =\left\vert 0\right\rangle \text{, }%
A_{0}\left\vert 1\right\rangle =\sqrt{1-\gamma}\left\vert 1\right\rangle
\text{, }%
\end{equation}
and,%
\begin{equation}
A_{1}\left\vert 0\right\rangle \equiv0\text{, }A_{1}\left\vert 1\right\rangle
=\sqrt{\gamma}\left\vert 0\right\rangle \text{,}%
\end{equation}
respectively. The codewords of the Leung et \textit{al}. $\left[  \left[
4\text{, }1\right]  \right]  $ quantum code are given by \cite{leung},%
\begin{equation}
\left\vert 0_{L}\right\rangle \overset{\text{def}}{=}\frac{1}{\sqrt{2}}\left(
\left\vert 0000\right\rangle +\left\vert 1111\right\rangle \right)  \text{
and, }\left\vert 1_{L}\right\rangle \overset{\text{def}}{=}\frac{1}{\sqrt{2}%
}\left(  \left\vert 0011\right\rangle +\left\vert 1100\right\rangle \right)
\text{.} \label{zuppa}%
\end{equation}
We underline that this code is a two-dimensional subspace of the
$16$-dimensional \emph{complex} Hilbert space $\mathcal{H}_{2}^{4}$ and is
spanned by self-complementary codewords. Recall that a code $\mathcal{C}$ is
called self-complementary if its codespace is spanned by codewords $\left\{
\left\vert c_{a}\right\rangle \right\}  $ defined as,%
\begin{equation}
\left\vert c_{a}\right\rangle \overset{\text{def}}{=}\frac{\left\vert
a\right\rangle +\left\vert \bar{a}\right\rangle }{\sqrt{2}}\text{,}%
\end{equation}
where $a$ is a binary string of length $n$ and $\bar{a}\overset{\text{def}}%
{=}\mathbf{1\oplus}a$ is the complement of $a$. In Appendix B, we show that,
in addition to the Leung et \textit{al}. four-qubit code, there are only two
additional two-dimensional subspaces spanned by self-complementary codewords
in $\mathcal{H}_{2}^{4}$ capable of error-correcting single AD-errors.

After the encoding operation, the total set of enlarged error operators is
given by the following $16$ enlarged error operators,
\begin{equation}
\mathcal{A}_{\text{total}}\overset{\text{def}}{=}\left\{
\begin{array}
[c]{c}%
A_{0000}\text{, }A_{1000}\text{, }A_{0100}\text{, }A_{0010}\text{, }%
A_{0001}\text{, }A_{1100}\text{, }A_{1010}\text{, }A_{1001}\text{, }%
A_{0110}\text{, }A_{0101}\text{, }A_{0011}\text{,}\\
\\
A_{1110}\text{, }A_{1011}\text{, }A_{0111}\text{, }A_{1101}\text{, }A_{1111}%
\end{array}
\right\}  \text{,}%
\end{equation}
where,%
\begin{equation}
16=2^{4}=\sum_{k=0}^{4}\binom{4}{k}=\binom{4}{0}+\binom{4}{1}+\binom{4}%
{2}+\binom{4}{3}+\binom{4}{4}\text{.}%
\end{equation}
Consider the following quantum state $\left\vert \psi\right\rangle $,%
\begin{equation}
\left\vert \psi\right\rangle \overset{\text{def}}{=}\alpha\left\vert
0_{L}\right\rangle +\beta\left\vert 1_{L}\right\rangle \text{,}%
\end{equation}
where $\alpha$, $\beta\in%
\mathbb{C}
$ and $\left\vert \alpha\right\vert ^{2}+\left\vert \beta\right\vert ^{2}=1$.
Then, the action of the weight-$0$ enlarged error operator $A_{0000}$ on
$\left\vert \psi\right\rangle $ reads,%
\begin{equation}
A_{0000}\left\vert \psi\right\rangle =\alpha\left[  \frac{\left\vert
0000\right\rangle +\left(  1-\gamma\right)  ^{2}\left\vert 1111\right\rangle
}{\sqrt{2}}\right]  +\beta\left[  \left(  1-\gamma\right)  \frac{\left(
\left\vert 0011\right\rangle +\left\vert 1100\right\rangle \right)  }{\sqrt
{2}}\right]  \text{.}%
\end{equation}
The action of the four weight-$1$ enlarged \ error operators is given by,%
\begin{align}
A_{1000}\left\vert \psi\right\rangle  &  =\sqrt{\frac{\gamma\left(
1-\gamma\right)  }{2}}\left[  \alpha\left(  1-\gamma\right)  \left\vert
0111\right\rangle +\beta\left\vert 0100\right\rangle \right]  \text{,
}A_{0100}\left\vert \psi\right\rangle =\sqrt{\frac{\gamma\left(
1-\gamma\right)  }{2}}\left[  \alpha\left(  1-\gamma\right)  \left\vert
1011\right\rangle +\beta\left\vert 1000\right\rangle \right]  \text{,}%
\nonumber\\
A_{0010}\left\vert \psi\right\rangle  &  =\sqrt{\frac{\gamma\left(
1-\gamma\right)  }{2}}\left[  \alpha\left(  1-\gamma\right)  \left\vert
1101\right\rangle +\beta\left\vert 0001\right\rangle \right]  \text{,
}A_{0001}\left\vert \psi\right\rangle =\sqrt{\frac{\gamma\left(
1-\gamma\right)  }{2}}\left[  \alpha\left(  1-\gamma\right)  \left\vert
1110\right\rangle +\beta\left\vert 0010\right\rangle \right]  \text{.}%
\end{align}
The action of the six weight-$2$ enlarged error operators reads,%
\begin{align}
A_{1100}\left\vert \psi\right\rangle  &  =\frac{\gamma}{\sqrt{2}}\left[
\alpha\left(  1-\gamma\right)  \left\vert 0011\right\rangle +\beta\left\vert
0000\right\rangle \right]  \text{, }A_{1010}\left\vert \psi\right\rangle
=\frac{\alpha}{\sqrt{2}}\gamma\left(  1-\gamma\right)  \left\vert
0101\right\rangle \text{, }\nonumber\\
A_{1001}\left\vert \psi\right\rangle  &  =\frac{\alpha}{\sqrt{2}}\gamma\left(
1-\gamma\right)  \left\vert 0110\right\rangle \text{, }A_{0110}\left\vert
\psi\right\rangle =\frac{\alpha}{\sqrt{2}}\gamma\left(  1-\gamma\right)
\left\vert 1001\right\rangle \text{, }\nonumber\\
A_{0101}\left\vert \psi\right\rangle  &  =\frac{\alpha}{\sqrt{2}}\gamma\left(
1-\gamma\right)  \left\vert 1010\right\rangle \text{, }A_{0011}\left\vert
\psi\right\rangle =\frac{\gamma}{\sqrt{2}}\left[  \alpha\left(  1-\gamma
\right)  \left\vert 1100\right\rangle +\beta\left\vert 0000\right\rangle
\right]  \text{.}%
\end{align}
The action of the four weight-$3$ enlarged \ error operators is given by,%
\begin{align}
A_{1110}\left\vert \psi\right\rangle  &  =\frac{\alpha}{\sqrt{2}}\gamma
^{\frac{3}{2}}\sqrt{1-\gamma}\left\vert 0001\right\rangle \text{, }%
A_{1011}\left\vert \psi\right\rangle =\frac{\alpha}{\sqrt{2}}\gamma^{\frac
{3}{2}}\sqrt{1-\gamma}\left\vert 0100\right\rangle \text{, }\nonumber\\
A_{0111}\left\vert \psi\right\rangle  &  =\frac{\alpha}{\sqrt{2}}\gamma
^{\frac{3}{2}}\sqrt{1-\gamma}\left\vert 1000\right\rangle \text{, }%
A_{1101}\left\vert \psi\right\rangle =\frac{\alpha}{\sqrt{2}}\gamma^{\frac
{3}{2}}\sqrt{1-\gamma}\left\vert 0010\right\rangle \text{.}%
\end{align}
Finally, the action of the weight-$4$ enlarged \ error operator reads,%
\begin{equation}
A_{1111}\left\vert \psi\right\rangle =\frac{\alpha}{\sqrt{2}}\gamma
^{2}\left\vert 0000\right\rangle \text{.}%
\end{equation}
For the sake of completeness, observe that%
\begin{align}
\sum_{i_{1}\text{, }i_{2}\text{, }i_{3}\text{, }i_{4}=0}^{1}P\left(
A_{i_{1}i_{2}i_{3}i_{4}}\right)   &  =\sum_{i_{1}\text{, }i_{2}\text{, }%
i_{3}\text{, }i_{4}=0}^{1}\text{Tr}\left(  A_{i_{1}i_{2}i_{3}i_{4}}\left\vert
\psi\right\rangle \left\langle \psi\right\vert A_{i_{1}i_{2}i_{3}i_{4}%
}^{\dagger}\right)  =\sum_{i_{1}\text{, }i_{2}\text{, }i_{3}\text{, }i_{4}%
=0}^{1}\left\langle \psi\left\vert A_{i_{1}i_{2}i_{3}i_{4}}^{\dagger}%
A_{i_{1}i_{2}i_{3}i_{4}}\right\vert \psi\right\rangle \nonumber\\
& \nonumber\\
&  =\left(  \frac{\alpha^{2}}{2}+\frac{\left(  1-\gamma\right)  ^{4}}{2}%
\alpha^{2}+\beta^{2}\left(  1-\gamma\right)  ^{2}\right)  +4\left(
\frac{\gamma\left(  1-\gamma\right)  }{2}\left(  \alpha^{2}\left(
1-\gamma\right)  ^{2}+\beta^{2}\right)  \right)  +\nonumber\\
& \nonumber\\
&  +\gamma^{2}\left(  \alpha^{2}\left(  1-\gamma\right)  ^{2}+\beta
^{2}\right)  +2\alpha^{2}\gamma^{2}\left(  1-\gamma\right)  ^{2}+2\alpha
^{2}\gamma^{3}\left(  1-\gamma\right)  +\frac{\alpha^{2}}{2}\gamma
^{4}\nonumber\\
& \nonumber\\
&  =\alpha^{2}+\beta^{2}\equiv\left\vert \alpha\right\vert ^{2}+\left\vert
\beta\right\vert ^{2}=1\text{,}%
\end{align}
where $P\left(  A_{i_{1}i_{2}i_{3}i_{4}}\right)  $ denotes the probability
that $A_{i_{1}i_{2}i_{3}i_{4}}$ occurs. We recall that\emph{\ }for a given
code $\mathcal{C}$, the set of detectable errors is closed under linear
combinations. That is, if $E_{1}$ and $E_{2}$ are both detectable, then so is
$\alpha E_{1}+\beta E_{2}$. This useful property implies that to check
detectability, one has to consider only the elements of a linear basis for the
space of errors of interest. An enlarged error operator $A_{k}$ is detectable
if and only if,%
\begin{equation}
P_{\mathcal{C}}A_{k}P_{\mathcal{C}}\propto P_{\mathcal{C}}\text{,} \label{con}%
\end{equation}
where $P_{\mathcal{C}}\overset{\text{def}}{=}\left\vert 0_{L}\right\rangle
\left\langle 0_{L}\right\vert +\left\vert 1_{L}\right\rangle \left\langle
1_{L}\right\vert $ is the projector on the codespace $\mathcal{C}$. Condition
(\ref{con}) requires that for detectable errors it must be,%
\begin{equation}
\left\langle 0_{L}\left\vert A_{k}\right\vert 0_{L}\right\rangle =\left\langle
1_{L}\left\vert A_{k}\right\vert 1_{L}\right\rangle \text{ and, }\left\langle
0_{L}\left\vert A_{k}\right\vert 1_{L}\right\rangle =0=\left\langle
1_{L}\left\vert A_{k}\right\vert 0_{L}\right\rangle \text{.}%
\end{equation}
For the weight-$0$ enlarged error operator $A_{0000}$, we have%
\begin{align}
\left\langle 0_{L}\left\vert A_{0000}\right\vert 0_{L}\right\rangle  &
=1-\gamma+\frac{1}{2}\gamma^{2}=1-\gamma+\mathcal{O}\left(  \gamma^{2}\right)
\text{, }\left\langle 1_{L}\left\vert A_{0000}\right\vert 1_{L}\right\rangle
=1-\gamma\text{,}\nonumber\\
& \nonumber\\
\left\langle 0_{L}\left\vert A_{0000}\right\vert 1_{L}\right\rangle  &
=0=\left\langle 1_{L}\left\vert A_{0000}\right\vert 0_{L}\right\rangle
\text{.}%
\end{align}
Therefore, $A_{0000}$ is detectable. Similarly, it turns out that the four
weight-$1$ enlarged \ error operators $A_{1000}$, $A_{0100}$, $A_{0010}$ and
$A_{0001}$ are detectable. For instance, for $A_{0100}$ we obtain%
\begin{equation}
\left\langle 0_{L}\left\vert A_{0100}\right\vert 0_{L}\right\rangle
=0=\left\langle 1_{L}\left\vert A_{0100}\right\vert 1_{L}\right\rangle \text{
and, }\left\langle 0_{L}\left\vert A_{0100}\right\vert 1_{L}\right\rangle
=0=\left\langle 1_{L}\left\vert A_{0100}\right\vert 0_{L}\right\rangle
\text{.}%
\end{equation}
Considering the weight-$2$ enlarged \ error operators, it follows that
$A_{1100}$ and $A_{0011}$ are not detectable since%
\begin{equation}
\left\langle 0_{L}\left\vert A_{1100}\right\vert 0_{L}\right\rangle
=0=\left\langle 1_{L}\left\vert A_{1100}\right\vert 1_{L}\right\rangle \text{
and, }\left\langle 0_{L}\left\vert A_{1100}\right\vert 1_{L}\right\rangle
=\frac{\gamma}{2}\neq0\text{, }\left\langle 1_{L}\left\vert A_{1100}%
\right\vert 0_{L}\right\rangle =\frac{\gamma\left(  1-\gamma\right)  }{2}%
\neq0\text{,}%
\end{equation}
and,%
\begin{equation}
\left\langle 0_{L}\left\vert A_{0011}\right\vert 0_{L}\right\rangle
=0=\left\langle 1_{L}\left\vert A_{0011}\right\vert 1_{L}\right\rangle \text{
and, }\left\langle 0_{L}\left\vert A_{0011}\right\vert 1_{L}\right\rangle
=\frac{\gamma}{2}\neq0\text{, }\left\langle 1_{L}\left\vert A_{0011}%
\right\vert 0_{L}\right\rangle =\frac{\gamma\left(  1-\gamma\right)  }{2}%
\neq0\text{.}%
\end{equation}
On the contrary the weight-$2$ enlarged error operators $A_{1010}$, $A_{1001}%
$, $A_{0110}$ and $A_{0101}$ are detectable. For instance, for $A_{1010}$ we
have%
\begin{equation}
\left\langle 0_{L}\left\vert A_{1010}\right\vert 0_{L}\right\rangle
=0=\left\langle 1_{L}\left\vert A_{1010}\right\vert 1_{L}\right\rangle \text{
and, }\left\langle 0_{L}\left\vert A_{1010}\right\vert 1_{L}\right\rangle
=0=\left\langle 1_{L}\left\vert A_{1010}\right\vert 0_{L}\right\rangle
\text{.}%
\end{equation}
The four weight-$3$ enlarged error operators $A_{1110}$, $A_{1011}$,
$A_{0111}$ and $A_{1101}$ are all detectable. For each one of them we get the
same type of relations which hold true for $A_{1110}$. For instance,%
\begin{equation}
\left\langle 0_{L}\left\vert A_{1110}\right\vert 0_{L}\right\rangle
=0=\left\langle 1_{L}\left\vert A_{1110}\right\vert 1_{L}\right\rangle \text{
and, }\left\langle 0_{L}\left\vert A_{1110}\right\vert 1_{L}\right\rangle
=0=\left\langle 1_{L}\left\vert A_{1110}\right\vert 0_{L}\right\rangle
\text{.}%
\end{equation}
Finally, the weight-$4$ enlarged \ error operator $A_{1111}$ is not detectable
since%
\begin{equation}
\left\langle 0_{L}\left\vert A_{1111}\right\vert 0_{L}\right\rangle
=\frac{\gamma^{2}}{2}\neq0=\left\langle 1_{L}\left\vert A_{1111}\right\vert
1_{L}\right\rangle \text{ and, }\left\langle 0_{L}\left\vert A_{1111}%
\right\vert 1_{L}\right\rangle =0=\left\langle 1_{L}\left\vert A_{1111}%
\right\vert 0_{L}\right\rangle \text{.}%
\end{equation}
We point out that the physical reason why $A_{0000}$ is detectable and
$A_{1111}$ is not detectable is as follows: there is a nonzero probability for
the error $A_{0000}$ to occur within the considered orders (up to linear
orders in gamma), whereas the error $A_{1111}$ would simply never occur in
those allowed orders ($0$-th and $1$-st in gamma); in other words, by ignoring
terms proportional to $\gamma^{2}$, the $A_{1111}$ error (four photons get
lost) simply does not exist and thus it cannot be referred to as detectable.
In conclusion, we have%
\begin{equation}
\mathcal{A}_{\text{detectable}}\overset{\text{def}}{=}\mathcal{A}%
_{\text{total}}/\left\{  A_{0011}\text{, }A_{1100}\text{, }A_{1111}\right\}
\subseteq\mathcal{A}_{\text{total}}\text{.}%
\end{equation}
We also recall that the notion of correctability depends on all the errors in
the set under consideration and, unlike detectability, cannot be applied to
individual errors \cite{zurek}. Furthermore, it is important to note that a
linear combination of correctable errors is also a correctable error. A set of
enlarged error operators $\left\{  A_{k}\right\}  $ is correctable iff,%
\begin{equation}
P_{\mathcal{C}}A_{l}^{\dagger}A_{m}P_{\mathcal{C}}\propto P_{\mathcal{C}%
}\text{.} \label{con2}%
\end{equation}
Condition (\ref{con2}) requires that for correctable errors it must be,%
\begin{equation}
\left\langle 0_{L}\left\vert A_{l}^{\dagger}A_{m}\right\vert 0_{L}%
\right\rangle =\left\langle 1_{L}\left\vert A_{l}^{\dagger}A_{m}\right\vert
1_{L}\right\rangle \text{ and, }\left\langle 0_{L}\left\vert A_{l}^{\dagger
}A_{m}\right\vert 1_{L}\right\rangle =0=\left\langle 1_{L}\left\vert
A_{l}^{\dagger}A_{m}\right\vert 0_{L}\right\rangle \text{.} \label{KL}%
\end{equation}
In the case under investigation, it turns out that the following set of
enlarged error operators is correctable%
\begin{equation}
\mathcal{A}_{\text{correctable}}\overset{\text{def}}{=}\left\{  A_{0000}%
\text{, }A_{1000}\text{, }A_{0100}\text{, }A_{0010}\text{, }A_{0001}\right\}
\subseteq\mathcal{A}_{\text{detectable}}\subseteq\mathcal{A}_{\text{total}%
}\text{.}%
\end{equation}
When $l\neq m$, error operators in $\mathcal{A}_{\text{correctable}}$
perfectly (arbitrary order in $\gamma$) satisfy the conditions in (\ref{KL}).
When $l=m$, we have%
\begin{align}
\left\langle 0_{L}\left\vert A_{0000}^{\dagger}A_{0000}\right\vert
1_{L}\right\rangle  &  =0=\left\langle 1_{L}\left\vert A_{0000}^{\dagger
}A_{0000}\right\vert 0_{L}\right\rangle \text{, }\left\langle 0_{L}\left\vert
A_{0000}^{\dagger}A_{0000}\right\vert 0_{L}\right\rangle =1-2\gamma
+\mathcal{O}\left(  \gamma^{2}\right)  =\left\langle 1_{L}\left\vert
A_{0000}^{\dagger}A_{0000}\right\vert 1_{L}\right\rangle \text{,}\nonumber\\
& \nonumber\\
\left\langle 0_{L}\left\vert A_{1000}^{\dagger}A_{1000}\right\vert
1_{L}\right\rangle  &  =0=\left\langle 1_{L}\left\vert A_{1000}^{\dagger
}A_{1000}\right\vert 0_{L}\right\rangle \text{, }\left\langle 0_{L}\left\vert
A_{1000}^{\dagger}A_{1000}\right\vert 0_{L}\right\rangle =\frac{\gamma}%
{2}+\mathcal{O}\left(  \gamma^{2}\right)  =\left\langle 1_{L}\left\vert
A_{1000}^{\dagger}A_{1000}\right\vert 1_{L}\right\rangle \text{,}\nonumber\\
& \nonumber\\
\left\langle 0_{L}\left\vert A_{0100}^{\dagger}A_{0100}\right\vert
1_{L}\right\rangle  &  =0=\left\langle 1_{L}\left\vert A_{0100}^{\dagger
}A_{0100}\right\vert 0_{L}\right\rangle \text{, }\left\langle 0_{L}\left\vert
A_{0100}^{\dagger}A_{0100}\right\vert 0_{L}\right\rangle =\frac{\gamma}%
{2}+\mathcal{O}\left(  \gamma^{2}\right)  =\left\langle 1_{L}\left\vert
A_{0100}^{\dagger}A_{0100}\right\vert 1_{L}\right\rangle \text{,}\nonumber\\
& \nonumber\\
\left\langle 0_{L}\left\vert A_{0010}^{\dagger}A_{0010}\right\vert
1_{L}\right\rangle  &  =0=\left\langle 1_{L}\left\vert A_{0010}^{\dagger
}A_{0010}\right\vert 0_{L}\right\rangle \text{, }\left\langle 0_{L}\left\vert
A_{0010}^{\dagger}A_{0010}\right\vert 0_{L}\right\rangle =\frac{\gamma}%
{2}+\mathcal{O}\left(  \gamma^{2}\right)  =\left\langle 1_{L}\left\vert
A_{0010}^{\dagger}A_{0010}\right\vert 1_{L}\right\rangle \text{,}\nonumber\\
& \nonumber\\
\left\langle 0_{L}\left\vert A_{0001}^{\dagger}A_{0001}\right\vert
1_{L}\right\rangle  &  =0=\left\langle 1_{L}\left\vert A_{0001}^{\dagger
}A_{0001}\right\vert 0_{L}\right\rangle \text{, }\left\langle 0_{L}\left\vert
A_{0001}^{\dagger}A_{0001}\right\vert 0_{L}\right\rangle =\frac{\gamma}%
{2}+\mathcal{O}\left(  \gamma^{2}\right)  =\left\langle 1_{L}\left\vert
A_{0001}^{\dagger}A_{0001}\right\vert 1_{L}\right\rangle \text{,}%
\end{align}
therefore, the KL-conditions in (\ref{KL}) are only approximately fulfilled to
the first order in the damping parameter $\gamma$.

\subsubsection{The standard QEC recovery operation}

Let us construct now the standard QEC recovery operators. Following our
remarks in Section II, we observe that the suitable orthonormal basis
$\mathcal{B}_{\mathcal{H}_{2}^{4}}$ for the $16$-dimensional \emph{complex}
Hilbert space $\mathcal{H}_{2}^{4}$ reads,%
\begin{equation}
\mathcal{B}_{\mathcal{H}_{2}^{4}}\overset{\text{def}}{=}\left\{  \left\vert
v_{r}^{j_{L}}\right\rangle \text{, }\left\vert o_{k}\right\rangle \right\}
\text{,}%
\end{equation}
with,%
\begin{align}
&  \left\vert v_{0}\right\rangle \overset{\text{def}}{=}\frac{\left\vert
0000\right\rangle +\left(  1-\gamma\right)  ^{2}\left\vert 1111\right\rangle
}{\sqrt{1+\left(  1-\gamma\right)  ^{4}}}\text{, }\left\vert v_{1}%
\right\rangle \overset{\text{def}}{=}\frac{\left\vert 0011\right\rangle
+\left\vert 1100\right\rangle }{\sqrt{2}}\text{, }\left\vert v_{2}%
\right\rangle \overset{\text{def}}{=}\left\vert 0111\right\rangle \text{,
}\left\vert v_{3}\right\rangle \overset{\text{def}}{=}\left\vert
0100\right\rangle \text{,}\nonumber\\
& \nonumber\\
&  \left\vert v_{4}\right\rangle \overset{\text{def}}{=}\left\vert
1011\right\rangle \text{, }\left\vert v_{5}\right\rangle \overset{\text{def}%
}{=}\left\vert 1000\right\rangle \text{, }\left\vert v_{6}\right\rangle
\overset{\text{def}}{=}\left\vert 1101\right\rangle \text{, }\left\vert
v_{7}\right\rangle \overset{\text{def}}{=}\left\vert 0001\right\rangle \text{,
}\left\vert v_{8}\right\rangle \overset{\text{def}}{=}\left\vert
1110\right\rangle \text{, }\nonumber\\
& \nonumber\\
&  \text{ }\left\vert v_{9}\right\rangle \overset{\text{def}}{=}\left\vert
0010\right\rangle \text{,}%
\end{align}
and,%
\begin{align}
\left\vert o_{1}\right\rangle  &  \equiv\left\vert v_{10}\right\rangle
\overset{\text{def}}{=}\left\vert 0101\right\rangle \text{, }\left\vert
o_{2}\right\rangle \equiv\left\vert v_{11}\right\rangle \overset{\text{def}%
}{=}\left\vert 0110\right\rangle \text{, }\left\vert o_{3}\right\rangle
\equiv\left\vert v_{12}\right\rangle \overset{\text{def}}{=}\left\vert
1001\right\rangle \text{, }\left\vert o_{4}\right\rangle \equiv\left\vert
v_{13}\right\rangle \overset{\text{def}}{=}\left\vert 1010\right\rangle
\text{,}\nonumber\\
& \nonumber\\
\left\vert o_{5}\right\rangle  &  \equiv\left\vert v_{14}\right\rangle
\overset{\text{def}}{=}\frac{\left(  1-\gamma\right)  ^{2}\left\vert
0000\right\rangle -\left\vert 1111\right\rangle }{\sqrt{1+\left(
1-\gamma\right)  ^{4}}}\text{, }\left\vert o_{6}\right\rangle \equiv\left\vert
v_{15}\right\rangle \overset{\text{def}}{=}\frac{\left\vert 0011\right\rangle
-\left\vert 1100\right\rangle }{\sqrt{2}}\text{.}%
\end{align}
The standard QEC recovery superoperator $\mathcal{R}$ is given by,%
\begin{equation}
\mathcal{R}\overset{\text{def}}{=}\left\{  R_{0}\text{, }R_{1}\text{, }%
R_{2}\text{, }R_{3}\text{, }R_{4}\text{, }\hat{O}\right\}  \text{,}%
\end{equation}
where,%
\begin{align}
&  R_{0}\overset{\text{def}}{=}\left\vert 0_{L}\right\rangle \left\langle
v_{0}\right\vert +\left\vert 1_{L}\right\rangle \left\langle v_{1}\right\vert
\text{, }\text{ }R_{1}\overset{\text{def}}{=}\left\vert 0_{L}\right\rangle
\left\langle v_{2}\right\vert +\left\vert 1_{L}\right\rangle \left\langle
v_{3}\right\vert \text{, }\text{ }R_{2}\overset{\text{def}}{=}\left\vert
0_{L}\right\rangle \left\langle v_{4}\right\vert +\left\vert 1_{L}%
\right\rangle \left\langle v_{5}\right\vert \text{,}\nonumber\\
& \nonumber\\
&  \text{ }R_{3}\overset{\text{def}}{=}\left\vert 0_{L}\right\rangle
\left\langle v_{6}\right\vert +\left\vert 1_{L}\right\rangle \left\langle
v_{7}\right\vert \text{, }\text{ }R_{4}\overset{\text{def}}{=}\left\vert
0_{L}\right\rangle \left\langle v_{8}\right\vert +\left\vert 1_{L}%
\right\rangle \left\langle v_{9}\right\vert \text{, }\text{ }\hat{O}%
\overset{\text{def}}{=}\sum_{k=1}^{6}\left\vert o_{k}\right\rangle
\left\langle o_{k}\right\vert \text{.} \label{99}%
\end{align}
We remark that, unlike the exact case, we have now:

\begin{itemize}
\item The two eigenvalues $\lambda_{\max}$ and $\lambda_{\min}$ of the
$\left(  2\times2\right)  $-matrix associated with the operators
$P_{\mathcal{C}}A_{l}^{\dagger}A_{m}P_{\mathcal{C}}$ (with $A_{k}$ correctable
errors) on the codespace $\mathcal{C}$ do not coincide. For example, for
$P_{\mathcal{C}}A_{0}^{\dagger}A_{0}P_{\mathcal{C}}$ we have $\lambda_{\max
}=\frac{1+\left(  1-\gamma\right)  ^{4}}{2}$ and $\lambda_{\min}=\left(
1-\gamma\right)  ^{2}$. The discrepancy between the two eigenvalues is a
fingerprint of the non-unitarity of $A_{k}P_{\mathcal{C}}$ where $A_{k}$ is correctable;

\item The non-tracelessness of the operators $P_{\mathcal{C}}A_{l}^{\dagger
}A_{m}P_{\mathcal{C}}$ with $l\neq m$ is an indicator of the non-orthogonality
between $A_{m}P_{\mathcal{C}}$ and $A_{l}P_{\mathcal{C}}$;

\item The projector on the codespace $P_{\mathcal{C}}$ does not belong to
$\mathcal{R}$, the standard QEC recovery;

\item There exist recovery operators in $\mathcal{R}$ that are $\gamma
$-dependent, where $\gamma$ denotes the damping probability and is the single
parameter that characterizes the noise model being considered.
\end{itemize}

In this case, it turns out that the entanglement fidelity becomes,%
\begin{align}
&  \mathcal{F}_{\left[  \left[  4,1\right]  \right]  }^{\text{QEC-recovery}%
}\left(  \gamma\right)  \overset{\text{def}}{=}\frac{1}{\left(  2\right)
^{2}}\sum_{l\text{, }k}\left\vert \text{Tr}\left(  R_{k}A_{l}\right)
_{\mathcal{C}}\right\vert ^{2}\nonumber\\
& \nonumber\\
&  =\frac{1}{4}\left(  \sqrt{\frac{1+\left(  1-\gamma\right)  ^{4}}{2}}%
+\sqrt{\frac{2\left(  1-\gamma\right)  ^{2}}{2}}\right)  ^{2}+\left(
\sqrt{\frac{\gamma\left(  1-\gamma\right)  ^{3}}{2}}+\sqrt{\frac{\gamma\left(
1-\gamma\right)  }{2}}\right)  ^{2}+\left(  \frac{1}{4}\frac{2}{1+\left(
1-\gamma\right)  ^{4}}\left(  \frac{\gamma^{2}}{2}\right)  ^{2}\right)
+\nonumber\\
& \nonumber\\
&  +\left(  \frac{1}{4}\left(  \frac{\gamma^{2}\left(  1-\gamma\right)
^{2}\left(  \left(  1-\gamma\right)  ^{2}-1\right)  }{2\left(  1+\left(
1-\gamma\right)  ^{4}\right)  }\right)  ^{2}\right) \nonumber\\
& \nonumber\\
&  \approx1-2\gamma^{2}+O\left(  \gamma^{3}\right)  \text{,}%
\end{align}
that is,%
\begin{equation}
\mathcal{F}_{\left[  \left[  4,1\right]  \right]  }^{\text{QEC-recovery}%
}\left(  \gamma\right)  \approx1-2\gamma^{2}+O\left(  \gamma^{3}\right)
\text{.} \label{QEC-recovery}%
\end{equation}
We stress that $\mathcal{F}_{\left[  \left[  4,1\right]  \right]
}^{\text{QEC-recovery}}\left(  \gamma\right)  $ is, in principle, the sum of
$5\times16=80$-terms that arise by considering all the possible pairs $\left(
k\text{, }l\right)  $ with $k\in\left\{  0\text{,..., }4\right\}  $ and
$l\in\left\{  0\text{,..., }16\right\}  $. However, it turns out that only
$6$-terms are nonvanishing and contribute to the computation of the
entanglement fidelity. They are $\left\{  \left(  k\text{, }l\right)
\right\}  =\left\{  \left(  0\text{, }0\right)  \text{, }\left(  1\text{,
}1\right)  \text{, }\left(  2\text{, }2\right)  \text{, }\left(  3\text{,
}3\right)  \text{, }\left(  4\text{, }4\right)  \text{, }\left(  0\text{,
}15\right)  \right\}  $. Thus, not only the five correctable errors $\left\{
A_{0}\text{, }A_{1}\text{, }A_{2}\text{, }A_{3}\text{, }A_{4}\right\}  $ are
partially recoverable since%
\begin{equation}
\left\langle 0_{L}\left\vert A_{j}^{\prime\dagger}A_{j}^{\prime}\right\vert
0_{L}\right\rangle \neq\left\langle 1_{L}\left\vert A_{j}^{\prime\dagger}%
A_{j}^{\prime}\right\vert 1_{L}\right\rangle \text{ (for arbitrary orders in
}\gamma\text{),}%
\end{equation}
but there is also the emergence of an off-diagonal contribution $\left(
0\text{, }15\right)  $. Thus, unlike the exact scenario, we have for the
approximate case that:

\begin{itemize}
\item Not only the correctable errors are recoverable. Indeed, they are not
fully recoverable. Off-diagonal contributions do arise.
\end{itemize}

\subsubsection{The code-projected recovery operation}

As we have noticed in Eq. (\ref{99}), the standard QEC recovery does not
contain the projector on the codespace as possible recovery operator. In this
new case, the chosen (orthonormal) basis vectors spanning $\mathcal{H}_{2}%
^{4}$ are given by,%
\begin{align}
&  \left\vert v_{0}\right\rangle \overset{\text{def}}{=}\frac{\left\vert
0000\right\rangle +\left\vert 1111\right\rangle }{\sqrt{2}}\text{, }\left\vert
v_{1}\right\rangle \overset{\text{def}}{=}\frac{\left\vert 0011\right\rangle
+\left\vert 1100\right\rangle }{\sqrt{2}}\text{, }\left\vert v_{2}%
\right\rangle \overset{\text{def}}{=}\frac{\left\vert 0000\right\rangle
-\left\vert 1111\right\rangle }{\sqrt{2}}\text{, }\left\vert v_{3}%
\right\rangle \overset{\text{def}}{=}\frac{\left\vert 0011\right\rangle
-\left\vert 1100\right\rangle }{\sqrt{2}}\text{,}\nonumber\\
& \nonumber\\
&  \left\vert v_{4}\right\rangle \overset{\text{def}}{=}\left\vert
0111\right\rangle \text{, }\left\vert v_{5}\right\rangle \overset{\text{def}%
}{=}\left\vert 0100\right\rangle \text{, }\left\vert v_{6}\right\rangle
\overset{\text{def}}{=}\left\vert 1011\right\rangle \text{, }\left\vert
v_{7}\right\rangle \overset{\text{def}}{=}\left\vert 1000\right\rangle \text{,
}\left\vert v_{8}\right\rangle \overset{\text{def}}{=}\left\vert
1101\right\rangle \text{, }\left\vert v_{9}\right\rangle \overset{\text{def}%
}{=}\left\vert 0001\right\rangle \text{,}\nonumber\\
& \nonumber\\
&  \left\vert v_{10}\right\rangle \overset{\text{def}}{=}\left\vert
1110\right\rangle \text{, }\left\vert v_{11}\right\rangle \overset{\text{def}%
}{=}\left\vert 0010\right\rangle \text{, }\left\vert v_{12}\right\rangle
\overset{\text{def}}{=}\left\vert 1001\right\rangle \text{, }\left\vert
v_{13}\right\rangle \overset{\text{def}}{=}\left\vert 1010\right\rangle
\text{, }\left\vert v_{14}\right\rangle \overset{\text{def}}{=}\left\vert
0101\right\rangle \text{, }\left\vert v_{15}\right\rangle \overset{\text{def}%
}{=}\left\vert 0110\right\rangle \text{.}%
\end{align}
The code-projected recovery (CP recovery) becomes,%
\begin{equation}
\mathcal{R}\overset{\text{def}}{=}\left\{  R_{1}\text{, }R_{2}\text{, }%
R_{3}\text{, }R_{4}\text{, }R_{5}\text{, }R_{6}\text{, }R_{7}\text{, }%
R_{8}\text{, }R_{9}\text{, }R_{10}\right\}  \text{,} \label{fff}%
\end{equation}
where,%
\begin{align}
&  R_{1}\overset{\text{def}}{=}\left\vert 0_{L}\right\rangle \left\langle
0_{L}\right\vert +\left\vert 1_{L}\right\rangle \left\langle 1_{L}\right\vert
\text{,}\nonumber\\
& \nonumber\\
&  \text{ }R_{2}\overset{\text{def}}{=}\left\vert 0_{L}\right\rangle \left(
\frac{1}{\sqrt{2}}\left\langle 0000\right\vert -\frac{1}{\sqrt{2}}\left\langle
1111\right\vert \right)  +\left\vert 1_{L}\right\rangle \left(  \frac{1}%
{\sqrt{2}}\left\langle 0011\right\vert -\frac{1}{\sqrt{2}}\left\langle
1100\right\vert \right)  \text{,}\nonumber\\
& \nonumber\\
R_{3}\overset{\text{def}}{=}\left\vert 0_{L}\right\rangle \left\langle
0111\right\vert +\left\vert 1_{L}\right\rangle \left\langle 0100\right\vert
&  =R_{A_{1000}}\text{, }R_{4}\overset{\text{def}}{=}\left\vert 0_{L}%
\right\rangle \left\langle 1011\right\vert +\left\vert 1_{L}\right\rangle
\left\langle 1000\right\vert =R_{A_{_{0100}}}\text{, }R_{5}\overset
{\text{def}}{=}\left\vert 0_{L}\right\rangle \left\langle 1101\right\vert
+\left\vert 1_{L}\right\rangle \left\langle 0001\right\vert =R_{A_{0010}%
}\text{, }\nonumber\\
& \nonumber\\
R_{6}\overset{\text{def}}{=}\left\vert 0_{L}\right\rangle \left\langle
1110\right\vert +\left\vert 1_{L}\right\rangle \left\langle 0010\right\vert
&  =R_{A_{0001}}\text{, }R_{7}\overset{\text{def}}{=}\left\vert 0_{L}%
\right\rangle \left\langle 1001\right\vert =R_{A_{0110}}\text{, }R_{8}%
\overset{\text{def}}{=}\left\vert 0_{L}\right\rangle \left\langle
1010\right\vert =R_{A_{0101}}\text{, }\nonumber\\
& \nonumber\\
R_{9}\overset{\text{def}}{=}\left\vert 0_{L}\right\rangle \left\langle
0101\right\vert  &  =R_{A_{1010}}\text{, }R_{10}\overset{\text{def}}%
{=}\left\vert 0_{L}\right\rangle \left\langle 0110\right\vert =R_{A_{1001}%
}\text{.}%
\end{align}
Observe that,%
\begin{align}
\sum_{k=1}^{10}P_{k}\overset{\text{def}}{=}\sum_{k=1}^{10}R_{k}^{\dagger
}R_{k}  &  =R_{1}^{\dagger}R_{1}+R_{2}^{\dagger}R_{2}+R_{3}^{\dagger}%
R_{3}+R_{4}^{\dagger}R_{4}+R_{5}^{\dagger}R_{5}+R_{6}^{\dagger}R_{6}%
+R_{7}^{\dagger}R_{7}+R_{8}^{\dagger}R_{8}+R_{9}^{\dagger}R_{9}+R_{10}%
^{\dagger}R_{10}\nonumber\\
& \nonumber\\
&  =\left\vert 0000\right\rangle \left\langle 0000\right\vert +\text{...}%
+\left\vert 1111\right\rangle \left\langle 1111\right\vert =I_{2^{4}%
\times2^{4}}\text{.}%
\end{align}
The entanglement fidelity becomes,%
\begin{equation}
\mathcal{F}_{\left[  \left[  4,1\right]  \right]  }^{\text{CP-recovery}%
}\left(  \gamma\right)  \overset{\text{def}}{=}\frac{1}{\left(  2\right)
^{2}}\sum_{k=0}^{15}\sum_{l=1}^{10}\left\vert \text{Tr}\left(  R_{l}%
A_{k}^{\prime}\right)  _{\left\vert \mathcal{C}\right.  }\right\vert
^{2}\text{,} \label{108}%
\end{equation}
where $R_{l}\in\mathcal{R}$ and $A_{0}^{\prime}\overset{\text{def}}{=}%
A_{0000}$, $A_{1}^{\prime}\overset{\text{def}}{=}A_{1000}$,..., $A_{15}%
^{\prime}=A_{1111}$. We point out that both the recovery operators $R_{1}$ and
$R_{2}$ contribute to the entanglement fidelity in Eq. (\ref{108}) since,%
\begin{align}
\left\langle 0_{L}\left\vert R_{1}A_{0000}\right\vert 0_{L}\right\rangle  &
=1-\gamma+\frac{\gamma^{2}}{2}\text{, }\left\langle 1_{L}\left\vert
R_{1}A_{0000}\right\vert 1_{L}\right\rangle =1-\gamma\text{, }\nonumber\\
\left\langle 0_{L}\left\vert R_{2}A_{0000}\right\vert 0_{L}\right\rangle  &
=\gamma-\frac{\gamma^{2}}{2}\text{, }\left\langle 1_{L}\left\vert
R_{2}A_{0000}\right\vert 1_{L}\right\rangle =0\text{,}%
\end{align}
and,%
\begin{align}
\left\langle 0_{L}\left\vert R_{1}A_{1111}\right\vert 0_{L}\right\rangle
+\left\langle 1_{L}\left\vert R_{1}A_{1111}\right\vert 1_{L}\right\rangle  &
=\frac{\gamma^{2}}{2}+0=\frac{\gamma^{2}}{2}\text{,}\nonumber\\
\left\langle 0_{L}\left\vert R_{2}A_{1111}\right\vert 0_{L}\right\rangle
+\left\langle 1_{L}\left\vert R_{2}A_{1111}\right\vert 1_{L}\right\rangle  &
=\frac{\gamma^{2}}{2}+0=\frac{\gamma^{2}}{2}\text{.}%
\end{align}
Furthermore, the contributions of recovery operators $R_{3}$, $R_{4}$, $R_{5}%
$, $R_{6}$ are given by,%
\begin{align}
\left\langle 0_{L}\left\vert R_{3}A_{1000}\right\vert 0_{L}\right\rangle
+\left\langle 1_{L}\left\vert R_{3}A_{1000}\right\vert 1_{L}\right\rangle  &
=\left(  1-\gamma\right)  \sqrt{\frac{\gamma\left(  1-\gamma\right)  }{2}%
}+\sqrt{\frac{\gamma\left(  1-\gamma\right)  }{2}}\text{,}\nonumber\\
\left\langle 0_{L}\left\vert R_{4}A_{0100}\right\vert 0_{L}\right\rangle
+\left\langle 1_{L}\left\vert R_{4}A_{0100}\right\vert 1_{L}\right\rangle  &
=\left(  1-\gamma\right)  \sqrt{\frac{\gamma\left(  1-\gamma\right)  }{2}%
}+\sqrt{\frac{\gamma\left(  1-\gamma\right)  }{2}}\text{,}%
\end{align}
and,%
\begin{align}
\left\langle 0_{L}\left\vert R_{5}A_{0010}\right\vert 0_{L}\right\rangle
+\left\langle 1_{L}\left\vert R_{5}A_{0010}\right\vert 1_{L}\right\rangle  &
=\left(  1-\gamma\right)  \sqrt{\frac{\gamma\left(  1-\gamma\right)  }{2}%
}+\sqrt{\frac{\gamma\left(  1-\gamma\right)  }{2}}\text{,}\nonumber\\
\left\langle 0_{L}\left\vert R_{6}A_{0001}\right\vert 0_{L}\right\rangle
+\left\langle 1_{L}\left\vert R_{6}A_{0001}\right\vert 1_{L}\right\rangle  &
=\left(  1-\gamma\right)  \sqrt{\frac{\gamma\left(  1-\gamma\right)  }{2}%
}+\sqrt{\frac{\gamma\left(  1-\gamma\right)  }{2}}\text{,}%
\end{align}
respectively. Finally, the contribution arising from the recovery operators
$R_{7}$, $R_{8}$, $R_{9}$, $R_{10}$ becomes transparent once we consider the
following relations,
\begin{align}
\left\langle 0_{L}\left\vert R_{7}A_{0110}\right\vert 0_{L}\right\rangle  &
=\frac{\gamma\left(  1-\gamma\right)  }{\sqrt{2}}\text{, }\left\langle
1_{L}\left\vert R_{7}A_{0110}\right\vert 1_{L}\right\rangle =0\text{,}%
\nonumber\\
& \nonumber\\
\left\langle 0_{L}\left\vert R_{8}A_{0101}\right\vert 0_{L}\right\rangle
+\left\langle 1_{L}\left\vert R_{8}A_{0101}\right\vert 1_{L}\right\rangle  &
=\frac{\gamma\left(  1-\gamma\right)  }{\sqrt{2}}+0=\frac{\gamma\left(
1-\gamma\right)  }{\sqrt{2}}\text{,}\nonumber\\
& \nonumber\\
\left\langle 0_{L}\left\vert R_{9}A_{1010}\right\vert 0_{L}\right\rangle
+\left\langle 1_{L}\left\vert R_{9}A_{1010}\right\vert 1_{L}\right\rangle  &
=\frac{\gamma\left(  1-\gamma\right)  }{\sqrt{2}}+0=\frac{\gamma\left(
1-\gamma\right)  }{\sqrt{2}}\text{,}\nonumber\\
& \nonumber\\
\left\langle 0_{L}\left\vert R_{10}A_{1001}\right\vert 0_{L}\right\rangle
+\left\langle 1_{L}\left\vert R_{10}A_{1001}\right\vert 1_{L}\right\rangle  &
=\frac{\gamma\left(  1-\gamma\right)  }{\sqrt{2}}+0=\frac{\gamma\left(
1-\gamma\right)  }{\sqrt{2}}\text{.}%
\end{align}
We notice that the six enlarged error operators $A_{1100}$, $A_{0011}$,
$A_{1110}$, $A_{1011}$, $A_{0111}$, $A_{1101}$ do not contribute to the
computation of the entanglement fidelity $\mathcal{F}_{\left[  \left[
4,1\right]  \right]  }^{\text{CP-recovery}}\left(  \gamma\right)  $. Finally,
we obtain%
\begin{align}
\mathcal{F}_{\left[  \left[  4,1\right]  \right]  }^{\text{CP-recovery}%
}\left(  \gamma\right)   &  =\frac{1}{4}\left\{
\begin{array}
[c]{c}%
\left[  \left(  1-\gamma+\frac{\gamma^{2}}{2}\right)  +\left(  1-\gamma
\right)  \right]  ^{2}+\left(  \gamma-\frac{\gamma^{2}}{2}\right)
^{2}+2\left(  \frac{\gamma^{2}}{2}\right)  ^{2}+4\left[  \left(
2-\gamma\right)  \sqrt{\frac{\gamma\left(  1-\gamma\right)  }{2}}\right]
^{2}+\\
\\
+4\left[  \frac{\gamma\left(  1-\gamma\right)  }{\sqrt{2}}\right]  ^{2}%
\end{array}
\right\}  \text{,}\nonumber\\
& \nonumber\\
&  \approx1-\frac{7}{4}\gamma^{2}+\mathcal{O}\left(  \gamma^{3}\right)
\text{,}%
\end{align}
that is,%
\begin{equation}
\mathcal{F}_{\left[  \left[  4,1\right]  \right]  }^{\text{CP-recovery}%
}\left(  \gamma\right)  \approx1-\frac{7}{4}\gamma^{2}+\mathcal{O}\left(
\gamma^{3}\right)  \text{.} \label{CP-recovery}%
\end{equation}
From Eqs. (\ref{QEC-recovery}) and (\ref{CP-recovery}), it turns out that as
far as the entanglement fidelity concerns, the CP recovery scheme is more
successful than the standard QEC recovery scheme.

\subsubsection{An analytically-optimized Fletcher's-type channel-adapted
recovery operation}

In addition to the standard QEC and CP recovery schemes, it is possible to
consider additional recovery schemes such as an analytically-optimized version
of a channel-adapted recovery scheme as proposed by Fletcher et \textit{al}.
in \cite{fletcher1}.

Consider the CP recovery scheme in Eq. (\ref{fff}) where, however, the
recovery operators $R_{1}$ and $R_{2}$ are defined as \cite{fletcher1},
\begin{align}
&  R_{1}\overset{\text{def}}{=}\left\vert 0_{L}\right\rangle \left(
a\left\langle 0000\right\vert +b\left\langle 1111\right\vert \right)
+\left\vert 1_{L}\right\rangle \left(  \frac{1}{\sqrt{2}}\left\langle
0011\right\vert +\frac{1}{\sqrt{2}}\left\langle 1100\right\vert \right)
\nonumber\\
& \nonumber\\
&  =\left\vert 0_{L}\right\rangle \left(  a\left\langle 0000\right\vert
+b\left\langle 1111\right\vert \right)  +\left\vert 1_{L}\right\rangle
\left\langle 1_{L}\right\vert \text{,}%
\end{align}
and,%
\begin{equation}
R_{2}\overset{\text{def}}{=}\left\vert 0_{L}\right\rangle \left(  b^{\ast
}\left\langle 0000\right\vert -a^{\ast}\left\langle 1111\right\vert \right)
+\left\vert 1_{L}\right\rangle \left(  \frac{1}{\sqrt{2}}\left\langle
0011\right\vert -\frac{1}{\sqrt{2}}\left\langle 1100\right\vert \right)
\text{,}%
\end{equation}
respectively, where $a$, $b\in%
\mathbb{C}
$ with $\left\vert a\right\vert ^{2}+\left\vert b\right\vert ^{2}=1$. The
remaining eight recovery operators are defined just as in Eq. (\ref{fff}). The
set of all ten recovery operators forms the Fletcher et \textit{al}. recovery
operation $\mathcal{R}_{\text{Fletcher}}$. Note that,%
\begin{equation}
R_{1}^{\dagger}=a^{\ast}\left\vert 0000\right\rangle \left\langle
0_{L}\right\vert +b^{\ast}\left\vert 1111\right\rangle \left\langle
0_{L}\right\vert +\left\vert 1_{L}\right\rangle \left\langle 1_{L}\right\vert
\text{,} \label{i}%
\end{equation}
and,%
\begin{equation}
R_{2}^{\dagger}=b\left\vert 0000\right\rangle \left\langle 0_{L}\right\vert
-a\left\vert 1111\right\rangle \left\langle 0_{L}\right\vert +\frac{1}%
{\sqrt{2}}\left\vert 0011\right\rangle \left\langle 1_{L}\right\vert -\frac
{1}{\sqrt{2}}\left\vert 1100\right\rangle \left\langle 1_{L}\right\vert
\text{.} \label{ii}%
\end{equation}
Using (\ref{i}) and (\ref{ii}), we get%
\begin{equation}
R_{1}^{\dagger}R_{1}+R_{2}^{\dagger}R_{2}=\left\vert 0000\right\rangle
\left\langle 0000\right\vert +\left\vert 0011\right\rangle \left\langle
0011\right\vert +\left\vert 1100\right\rangle \left\langle 1100\right\vert
+\left\vert 1111\right\rangle \left\langle 1111\right\vert \text{.}%
\end{equation}
It turns out that for the Fletcher et \textit{al}. recovery (F-recovery)
operations,%
\begin{equation}
\sum_{k}P_{k}=\sum_{k}R_{k}^{\dagger}R_{k}=\left\vert 0000\right\rangle
\left\langle 0000\right\vert +\text{...}+\left\vert 1111\right\rangle
\left\langle 1111\right\vert =I_{2^{4}\times2^{4}}\text{.}%
\end{equation}
In this case, the entanglement fidelity $\mathcal{F}_{\left[  \left[
4,1\right]  \right]  }^{\text{F-recovery}}$ reads,%
\begin{equation}
\mathcal{F}_{\left[  \left[  4,1\right]  \right]  }^{\text{F-recovery}%
}\overset{\text{def}}{=}\frac{1}{\left(  2\right)  ^{2}}\sum_{k=0}^{15}%
\sum_{l=1}^{10}\left\vert \text{Tr}\left(  R_{l}A_{k}^{\prime}\right)
_{\left\vert \mathcal{C}\right.  }\right\vert ^{2}\text{,}%
\end{equation}
where $R_{l}\in\mathcal{R}_{\text{Fletcher}}$ and $A_{0}^{\prime}%
\overset{\text{def}}{=}A_{0000}$, $A_{1}^{\prime}\overset{\text{def}}%
{=}A_{1000}$,..., $A_{15}^{\prime}=A_{1111}$. Notice that,%
\begin{align}
\left\langle 0_{L}\left\vert R_{1}A_{0000}\right\vert 0_{L}\right\rangle  &
=\frac{a+b\left(  1-\gamma\right)  ^{2}}{\sqrt{2}}\text{, }\left\langle
1_{L}\left\vert R_{1}A_{0000}\right\vert 1_{L}\right\rangle =1-\gamma
\text{,}\nonumber\\
\left\langle 0_{L}\left\vert R_{2}A_{0000}\right\vert 0_{L}\right\rangle  &
=\frac{b^{\ast}-a^{\ast}\left(  1-\gamma\right)  ^{2}}{\sqrt{2}}\text{,
}\left\langle 1_{L}\left\vert R_{2}A_{0000}\right\vert 1_{L}\right\rangle
=0\text{,}\nonumber\\
\left\langle 0_{L}\left\vert R_{1}A_{1111}\right\vert 0_{L}\right\rangle  &
=a\frac{\gamma^{2}}{\sqrt{2}}\text{, }\left\langle 1_{L}\left\vert
R_{1}A_{1111}\right\vert 1_{L}\right\rangle =0\text{,}\nonumber\\
\left\langle 0_{L}\left\vert R_{2}A_{1111}\right\vert 0_{L}\right\rangle  &
=b^{\ast}\frac{\gamma^{2}}{\sqrt{2}}\text{, }\left\langle 1_{L}\left\vert
R_{2}A_{1111}\right\vert 1_{L}\right\rangle =0\text{,}%
\end{align}
while the remaining terms are the same as obtained in the previous analysis
performed with the traditional QEC recovery scheme. Following the line of
reasoning provide in the former computations, we get $\mathcal{F}_{\left[
\left[  4,1\right]  \right]  }^{\text{F-recovery}}\left(  \gamma\right)
\equiv\mathcal{F}_{\left[  \left[  4,1\right]  \right]  }\left(  a\text{,
}b\text{, }\gamma\right)  $ with,%
\begin{equation}
\mathcal{F}_{\left[  \left[  4,1\right]  \right]  }\left(  a\text{, }b\text{,
}\gamma\right)  =\frac{1}{4}\left\{  \left\vert \frac{a+b\left(
1-\gamma\right)  ^{2}}{\sqrt{2}}+\left(  1-\gamma\right)  \right\vert
^{2}+\left\vert \frac{b^{\ast}-a^{\ast}\left(  1-\gamma\right)  ^{2}}{\sqrt
{2}}\right\vert ^{2}+2\gamma\left(  1-\gamma\right)  \left(  2-\gamma\right)
^{2}+2\gamma^{2}\left(  1-\gamma\right)  ^{2}+\frac{\gamma^{4}}{2}\right\}
\text{.}%
\end{equation}
We wish to maximize $\mathcal{F}_{\left[  \left[  4,1\right]  \right]
}\left(  a\text{, }b\text{, }\gamma\right)  $. The problem is to find $\bar
{a}$ and $\bar{b}$ (perhaps $\gamma$-dependent quantities) such that
$\mathcal{F}_{\left[  \left[  4,1\right]  \right]  }\left(  \bar{a}\text{,
}\bar{b}\text{, }\gamma\right)  $ denotes the searched maximum,%
\begin{equation}
\mathcal{F}_{\left[  \left[  4,1\right]  \right]  }\left(  \bar{a}\text{,
}\bar{b}\text{, }\gamma\right)  =\underset{\left\vert a\right\vert
^{2}+\left\vert b\right\vert ^{2}=1}{\max}\mathcal{F}\left(  a\text{,
}b\text{, }\gamma\right)  \text{.}%
\end{equation}
It can be shown that (for details, see Appendix C),%
\begin{equation}
\mathcal{F}_{\left[  \left[  4,1\right]  \right]  }\left(  \bar{a}\text{,
}\bar{b}\text{, }\gamma\right)  =1-\frac{3}{2}\gamma^{2}+\mathcal{O}\left(
\gamma^{3}\right)  \text{,}%
\end{equation}
with%
\begin{equation}
\bar{a}\left(  \gamma\right)  \overset{\text{def}}{=}\frac{1}{\sqrt{1+\left(
1-\gamma\right)  ^{4}}}\text{ and, }\bar{b}\left(  \gamma\right)
\overset{\text{def}}{=}\frac{\left(  1-\gamma\right)  ^{2}}{\sqrt{1+\left(
1-\gamma\right)  ^{4}}}\text{.}%
\end{equation}
Finally,
\begin{equation}
\mathcal{F}_{\left[  \left[  4,1\right]  \right]  }^{\text{F-recovery}}\left(
\gamma\right)  \approx1-\frac{3}{2}\gamma^{2}+\mathcal{O}\left(  \gamma
^{3}\right)  \text{.} \label{F-recovery}%
\end{equation}
From Eqs. (\ref{QEC-recovery}), (\ref{CP-recovery}) and (\ref{F-recovery}), it
turns out that as far as the entanglement fidelity concerns, the
analytically-optimized F-recovery scheme is better than both the standard QEC
and CP recovery schemes. The comparison of the three recovery schemes employed
can be visualized in Fig. $1$.

\begin{figure}[ptb]
\centering
\includegraphics[width=0.5\textwidth]{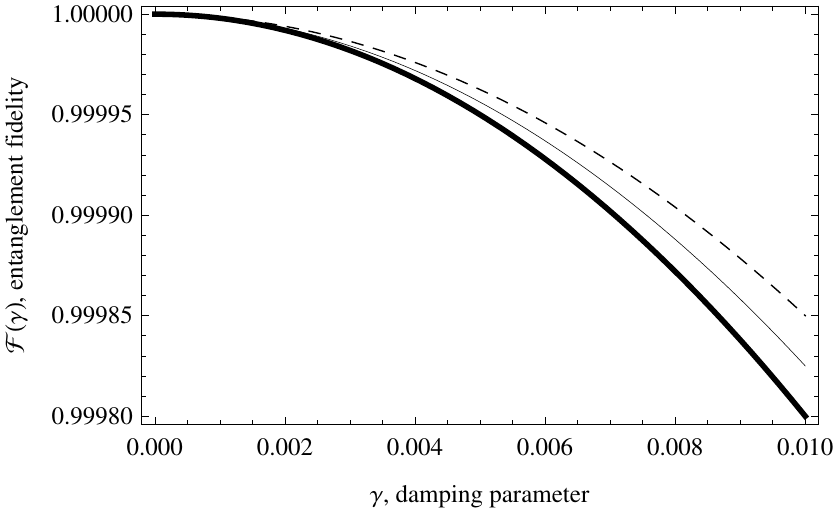}\caption{The truncated series
expansion of the entanglement fidelity $\mathcal{F}\left(  \gamma\right)  $
vs. the amplitude damping parameter $\gamma$ with $0\leq\gamma\leq10^{-2}$ for
the Leung et \textit{al.} four-qubit code for amplitude damping errors; the
Fletcher-type recovery (dashed line), the code-projected recovery (thin solid
line) and, the standard QEC-recovery (thick solid line).}%
\label{fig1}%
\end{figure}

\section{Final remarks}

In this article, we presented a comparative analysis of exact and approximate
quantum error correction by means of simple unabridged analytical
computations. For the sake of clarity, using primitive quantum codes, we
showed a detailed study of exact and approximate error correction for the two
simplest unital (Pauli errors) and nonunital (non-Pauli errors) noise models,
respectively. The similarities and differences between the two scenarios were
stressed. In addition, the performances of quantum codes quantified by means
of the entanglement fidelity for different recovery schemes were taken into
consideration in the approximate case.

Our main findings, some of which appear in the appendices to ease the
readability of the article, can be outlined as follows:

\medskip

\begin{enumerate}
\item We have explicitly constructed one of the recovery operators as
originally proposed by Leung et \textit{al}. in \cite{leung}. As a by-product,
we also found the correct version of Eq. $(41)$ in \cite{leung}. Our version
is represented by Eq. (\ref{correct}) in Appendix A.

\item We have explicitly discussed the similarities and differences between
exact and approximate-QEC schemes for very simple noise models and very common
stabilizer codes. Our analysis is purely analytical and no numerical
consideration is required. Thus, it is straightforward to follow and, we
believe, has considerable pedagogical and explanatory relevance. In
particular, the points to be stressed in the \emph{exact case} are:
\end{enumerate}

\begin{itemize}
\item The two eigenvalues $\lambda_{\max}$ and $\lambda_{\min}$ of the
$\left(  2\times2\right)  $-matrix associated with the operators
$P_{\mathcal{C}}A_{l}^{\dagger}A_{m}P_{\mathcal{C}}$ (with $A_{k}$ correctable
errors) on the codespace $\mathcal{C}$ coincide;

\item The projector on the codespace $P_{\mathcal{C}}$ belongs to the standard
QEC recovery $\mathcal{R}$;

\item All the recovery operators in $\mathcal{R}$ are $p$-independent;

\item The correctable errors are fully recoverable. No off-diagonal
contribution arises.
\end{itemize}

On the other side, the main points to be stressed in the \emph{approximate
case} are:

\begin{itemize}
\item The two eigenvalues $\lambda_{\max}$ and $\lambda_{\min}$ of the
$\left(  2\times2\right)  $-matrix associated with the operators
$P_{\mathcal{C}}A_{l}^{\dagger}A_{m}P_{\mathcal{C}}$ (with $A_{k}$ correctable
errors) on the codespace $\mathcal{C}$ do not coincide;

\item The discrepancy between the two eigenvalues is a fingerprint of the
non-unitarity of $A_{k}P_{\mathcal{C}}$ where $A_{k}$ is a correctable error;

\item The non-tracelessness of the operators $P_{\mathcal{C}}A_{l}^{\dagger
}A_{m}P_{\mathcal{C}}$ with $l\neq m$ is an indicator of the non-orthogonality
between $A_{m}P_{\mathcal{C}}$ and $A_{l}P_{\mathcal{C}}$;

\item The projector on the codespace $P_{\mathcal{C}}$ does not belong to the
standard QEC recovery $\mathcal{R}$;

\item There exist recovery operators in $\mathcal{R}$ that are $\gamma$-dependent;

\item The correctable errors are not fully recoverable. Off-diagonal
contributions do arise.
\end{itemize}

\begin{enumerate}
\item[3.] We have explicitly shown that there are only three possible
self-complementary quantum codes characterized by a two-dimensional subspace
of the sixteen-dimensional \emph{complex} Hilbert space $\mathcal{H}_{2}^{4}$
capable of error-correcting single-AD\ errors. Thus, in this regard, the Leung
et \textit{al}. four-qubit code is not unique. Our three codes appear in Eqs.
(\ref{a}), (\ref{b}) and (\ref{c}) in Appendix B.

\item[4.] In the approximate-QEC case, we have explicitly computed the
entanglement fidelity for three different recovery schemes. In particular, Eq.
(\ref{QEC-recovery}) for the standard QEC recovery has, to the best of our
knowledge, never appeared in the literature (neither numerically nor
analytically); furthermore, Eq. (\ref{CP-recovery}) for the code-projected
recovery is the analytical counterpart of the numerical finding presented in
\cite{fletcher1}; finally, Eq. (\ref{F-recovery}) represents our analytical
contribution to the understanding of the numerical result presented in
\cite{fletcher2}.
\end{enumerate}

\medskip

Although our investigation is limited to very simple noise models and very
simple codes, we hope that it will inspire other researchers to pursue novel
analytical studies of more realistic noise models and higher-dimensional
quantum codes. After all, for such type of investigations, analytical
computations can become considerably messy (as pointed out in \cite{isabel})
and understanding in an analytical fashion recovery maps numerically computed
can become quite a tricky task as well (as stressed in \cite{ng}).

\smallskip

In conclusion, also in view of these very last considerations, we are very
confident about the relevance of the pedagogical nature of our analytical
investigation carried out in this article.

\begin{acknowledgments}
We thank the ERA-Net CHIST-ERA project HIPERCOM for financial support.
\end{acknowledgments}

\appendix

\section{Unabridged computation of the recovery operation}

For the sake of clarity, we limit our analysis to the explicit computation of
the Leung et \textit{al}. recovery operator for the enlarged error operator
$A_{0000}$. In principle, the remaining recovery operators can be computed in
the same manner. Observe that in general we should be dealing with operators
acting on the $16$-dimensional \emph{complex} Hilbert space $\mathcal{H}%
_{2}^{4}$. However, in what follows, we shall take into consideration only
lower-dimensional matrix-representations of operators where the dimension is
limited to nontrivial contributions. For instance, for $A_{0000}$ and
$P_{\mathcal{C}}$, we consider their matrix-representation restricted to the
four-dimensional subspace of $\mathcal{H}_{2}^{4}$ spanned by the orthonormal
vectors,%
\begin{equation}
\left\{  \left\vert 0000\right\rangle \text{, }\left\vert 0011\right\rangle
\text{, }\left\vert 1100\right\rangle \text{, }\left\vert 1111\right\rangle
\right\}  \text{.}%
\end{equation}
We obtain,%
\begin{equation}
P_{\mathcal{C}}\overset{\text{def}}{=}\frac{1}{2}\left(
\begin{array}
[c]{cccc}%
1 & 0 & 0 & 1\\
0 & 1 & 1 & 0\\
0 & 1 & 1 & 0\\
1 & 0 & 0 & 1
\end{array}
\right)  \text{ and, }A_{0000}\overset{\text{def}}{=}\left(
\begin{array}
[c]{cccc}%
1 & 0 & 0 & 0\\
0 & 1-\gamma & 0 & 0\\
0 & 0 & 1-\gamma & 0\\
0 & 0 & 0 & \left(  1-\gamma\right)  ^{2}%
\end{array}
\right)  \text{.} \label{pa}%
\end{equation}
From Eq. (\ref{pa}), it follows that the two eigenvalues of $P_{\mathcal{C}%
}A_{0000}^{\dagger}A_{0000}P_{\mathcal{C}}$ are given by,%
\begin{equation}
\lambda_{\min}\overset{\text{def}}{=}\left(  1-\gamma\right)  ^{2}\text{ and,
}\lambda_{\max}\overset{\text{def}}{=}\frac{1+\left(  1-\gamma\right)  ^{4}%
}{2}\text{,}%
\end{equation}
while $\sqrt{P_{\mathcal{C}}A_{0000}^{\dagger}A_{0000}P_{\mathcal{C}}}$ reads,%
\begin{equation}
\sqrt{P_{\mathcal{C}}A_{0000}^{\dagger}A_{0000}P_{\mathcal{C}}}=\left(
\begin{array}
[c]{cccc}%
\frac{\left(  1-\gamma\right)  ^{4}+1}{4} & 0 & 0 & \frac{\left(
1-\gamma\right)  ^{4}+1}{4}\\
0 & \frac{\left(  1-\gamma\right)  ^{2}}{2} & \frac{\left(  1-\gamma\right)
^{2}}{2} & 0\\
0 & \frac{\left(  1-\gamma\right)  ^{2}}{2} & \frac{\left(  1-\gamma\right)
^{2}}{2} & 0\\
\frac{\left(  1-\gamma\right)  ^{4}+1}{4} & 0 & 0 & \frac{\left(
1-\gamma\right)  ^{4}+1}{4}%
\end{array}
\right)  ^{\frac{1}{2}}\text{.}%
\end{equation}
After some algebra, we have%
\begin{equation}
\sqrt{P_{\mathcal{C}}A_{0000}^{\dagger}A_{0000}P_{\mathcal{C}}}=\lambda
_{1}\left\vert v_{1}\right\rangle \left\langle v_{1}\right\vert +\lambda
_{2}\left\vert v_{2}\right\rangle \left\langle v_{2}\right\vert +\lambda
_{3}\left\vert v_{3}\right\rangle \left\langle v_{3}\right\vert +\lambda
_{4}\left\vert v_{4}\right\rangle \left\langle v_{4}\right\vert \text{,}
\label{cafe3}%
\end{equation}
where,%
\begin{equation}
\lambda_{1}\overset{\text{def}}{=}1-\gamma\text{, }\lambda_{2}\overset
{\text{def}}{=}0\text{, }\lambda_{3}\overset{\text{def}}{=}0\text{, }%
\lambda_{4}\overset{\text{def}}{=}\sqrt{\frac{1+\left(  1-\gamma\right)  ^{4}%
}{2}}\text{,} \label{cafe2}%
\end{equation}
and,%
\begin{equation}
\left\vert v_{1}\right\rangle \overset{\text{def}}{=}\frac{1}{\sqrt{2}}\left(
\begin{array}
[c]{c}%
0\\
1\\
1\\
0
\end{array}
\right)  \text{, }\left\vert v_{2}\right\rangle \overset{\text{def}}{=}%
\frac{1}{\sqrt{2}}\left(
\begin{array}
[c]{c}%
0\\
-1\\
1\\
0
\end{array}
\right)  \text{, }\left\vert v_{3}\right\rangle \overset{\text{def}}{=}%
\frac{1}{\sqrt{2}}\left(
\begin{array}
[c]{c}%
-1\\
0\\
0\\
1
\end{array}
\right)  \text{, }\left\vert v_{4}\right\rangle \overset{\text{def}}{=}%
\frac{1}{\sqrt{2}}\left(
\begin{array}
[c]{c}%
1\\
0\\
0\\
1
\end{array}
\right)  \text{.} \label{cafe1}%
\end{equation}
Substituting (\ref{cafe1}) and (\ref{cafe2}) into (\ref{cafe3}), we get%
\begin{equation}
\sqrt{P_{\mathcal{C}}A_{0000}^{\dagger}A_{0000}P_{\mathcal{C}}}=\frac{1}%
{2}\left(
\begin{array}
[c]{cccc}%
\sqrt{\frac{\left(  1-\gamma\right)  ^{4}+1}{2}} & 0 & 0 & \sqrt{\frac{\left(
1-\gamma\right)  ^{4}+1}{2}}\\
0 & 1-\gamma & 1-\gamma & 0\\
0 & 1-\gamma & 1-\gamma & 0\\
\sqrt{\frac{\left(  1-\gamma\right)  ^{4}+1}{2}} & 0 & 0 & \sqrt{\frac{\left(
1-\gamma\right)  ^{4}+1}{2}}%
\end{array}
\right)  \text{.}%
\end{equation}
Recall that the residue operator $\pi_{0000}$ is given by,%
\begin{equation}
\pi_{0000}\overset{\text{def}}{=}\sqrt{P_{\mathcal{C}}A_{0000}^{\dagger
}A_{0000}P_{\mathcal{C}}}-\sqrt{\lambda_{\min}}P_{C}\text{,}%
\end{equation}
that is,%
\begin{equation}
\pi_{0000}=\left(
\begin{array}
[c]{cccc}%
\frac{1}{2}\gamma+\frac{1}{2}\sqrt{\frac{1}{2}\left(  \gamma-1\right)
^{4}+\frac{1}{2}}-\frac{1}{2} & 0 & 0 & \frac{1}{2}\gamma+\frac{1}{2}%
\sqrt{\frac{1}{2}\left(  \gamma-1\right)  ^{4}+\frac{1}{2}}-\frac{1}{2}\\
0 & 0 & 0 & 0\\
0 & 0 & 0 & 0\\
\frac{1}{2}\gamma+\frac{1}{2}\sqrt{\frac{1}{2}\left(  \gamma-1\right)
^{4}+\frac{1}{2}}-\frac{1}{2} & 0 & 0 & \frac{1}{2}\gamma+\frac{1}{2}%
\sqrt{\frac{1}{2}\left(  \gamma-1\right)  ^{4}+\frac{1}{2}}-\frac{1}{2}%
\end{array}
\right)  \text{.} \label{correct}%
\end{equation}
Observe that for $\gamma\ll1$, $\pi_{0000}$ becomes%
\begin{equation}
\pi_{0000}=\frac{1}{2}\gamma^{2}\left(
\begin{array}
[c]{cccc}%
1 & 0 & 0 & 1\\
0 & 0 & 0 & 0\\
0 & 0 & 0 & 0\\
1 & 0 & 0 & 1
\end{array}
\right)  +\mathcal{O}\left(  \gamma^{4}\right)  \text{.}%
\end{equation}
Let us focus now on the computation of the unitary operator $U_{0000}$. From
Eqs. (\ref{pa}) and (\ref{cafe1}), we get%
\begin{align}
&  A_{0000}P_{\mathcal{C}}\left\vert v_{1}\right\rangle \overset{\text{def}%
}{=}\frac{1-\gamma}{\sqrt{2}}\left(
\begin{array}
[c]{c}%
0\\
1\\
1\\
0
\end{array}
\right)  \text{, }A_{0000}P_{\mathcal{C}}\left\vert v_{2}\right\rangle
\overset{\text{def}}{=}\left(
\begin{array}
[c]{c}%
0\\
0\\
0\\
0
\end{array}
\right)  \text{, }A_{0000}P_{\mathcal{C}}\left\vert v_{3}\right\rangle
\overset{\text{def}}{=}\left(
\begin{array}
[c]{c}%
0\\
0\\
0\\
0
\end{array}
\right)  \text{, }\nonumber\\
& \nonumber\\
&  A_{0000}P_{\mathcal{C}}\left\vert v_{4}\right\rangle \overset{\text{def}%
}{=}\text{ }\frac{1}{\sqrt{2}}\left(
\begin{array}
[c]{c}%
1\\
0\\
0\\
\left(  1-\gamma\right)  ^{2}%
\end{array}
\right)  \text{.}%
\end{align}
Consider the following basis given by,%
\begin{equation}
\left\{  \left\vert e_{1}\right\rangle \text{, }\left\vert e_{2}\right\rangle
\text{, }\left\vert e_{3}\right\rangle \text{, }\left\vert e_{4}\right\rangle
\right\}  \text{,}%
\end{equation}
with,%
\begin{align}
\left\vert e_{1}\right\rangle \overset{\text{def}}{=}\frac{A_{0000}%
P_{\mathcal{C}}\left\vert v_{1}\right\rangle }{\lambda_{1}}  &  =\frac
{1}{\sqrt{2}}\left(
\begin{array}
[c]{c}%
0\\
1\\
1\\
0
\end{array}
\right)  \text{, }\left\vert e_{2}\right\rangle \overset{\text{def}}{=}%
\frac{1}{\sqrt{2}}\left(
\begin{array}
[c]{c}%
0\\
-1\\
1\\
0
\end{array}
\right)  \text{, }\left\vert e_{3}\right\rangle \overset{\text{def}}{=}%
\frac{1}{\sqrt{2}}\left(
\begin{array}
[c]{c}%
-1\\
0\\
0\\
1
\end{array}
\right)  \text{, }\nonumber\\
& \nonumber\\
\left\vert e_{4}\right\rangle \overset{\text{def}}{=}\frac{A_{0000}%
P_{\mathcal{C}}\left\vert v_{4}\right\rangle }{\lambda_{4}}  &  =\frac
{1}{\sqrt{2}}\frac{1}{\sqrt{\frac{1+\left(  1-\gamma\right)  ^{4}}{2}}}\left(
\begin{array}
[c]{c}%
1\\
0\\
0\\
\left(  1-\gamma\right)  ^{2}%
\end{array}
\right)  \text{.}%
\end{align}
Applying the Gram-Schmidt orthonormalization procedure to $\left\{  \left\vert
e_{1}\right\rangle \text{, }\left\vert e_{2}\right\rangle \text{, }\left\vert
e_{3}\right\rangle \text{, }\left\vert e_{4}\right\rangle \right\}  $, we get%
\begin{align}
\left\vert E_{1}\right\rangle \overset{\text{def}}{=}\frac{A_{0000}%
P_{\mathcal{C}}\left\vert v_{1}\right\rangle }{\lambda_{1}}  &  =\frac
{1}{\sqrt{2}}\left(
\begin{array}
[c]{c}%
0\\
1\\
1\\
0
\end{array}
\right)  \text{, }\left\vert E_{2}\right\rangle \overset{\text{def}}{=}%
\frac{1}{\sqrt{2}}\left(
\begin{array}
[c]{c}%
0\\
-1\\
1\\
0
\end{array}
\right)  \text{, }\left\vert E_{3}\right\rangle \overset{\text{def}}{=}%
\frac{1}{\sqrt{2}}\frac{1}{\sqrt{\frac{1+\left(  1-\gamma\right)  ^{4}}{2}}%
}\left(
\begin{array}
[c]{c}%
-\left(  1-\gamma\right)  ^{2}\\
0\\
0\\
1
\end{array}
\right)  \text{, }\nonumber\\
& \nonumber\\
\left\vert E_{4}\right\rangle \overset{\text{def}}{=}\frac{A_{0000}%
P_{\mathcal{C}}\left\vert v_{4}\right\rangle }{\lambda_{4}}  &  =\frac
{1}{\sqrt{2}}\frac{1}{\sqrt{\frac{1+\left(  1-\gamma\right)  ^{4}}{2}}}\left(
\begin{array}
[c]{c}%
1\\
0\\
0\\
\left(  1-\gamma\right)  ^{2}%
\end{array}
\right)  \text{,} \label{E}%
\end{align}
with $\left\langle E_{l}\left\vert E_{k}\right.  \right\rangle =\delta_{lk}$.
Finally, the unitary operator $U_{0000}$ reads,%
\begin{equation}
U_{0000}\overset{\text{def}}{=}\left\vert E_{1}\right\rangle \left\langle
v_{1}\right\vert +\left\vert E_{2}\right\rangle \left\langle v_{2}\right\vert
+\left\vert E_{3}\right\rangle \left\langle v_{3}\right\vert +\left\vert
E_{4}\right\rangle \left\langle v_{4}\right\vert \text{,}%
\end{equation}
that is, using Eqs. (\ref{cafe1}) and (\ref{E}),%
\begin{equation}
U_{0000}\overset{\text{def}}{=}\left(
\begin{array}
[c]{cccc}%
\frac{1}{\sqrt{2}}\frac{1+\left(  1-\gamma\right)  ^{2}}{\sqrt{1+\left(
1-\gamma\right)  ^{4}}} & 0 & 0 & \frac{1}{\sqrt{2}}\frac{1-\left(
1-\gamma\right)  ^{2}}{\sqrt{1+\left(  1-\gamma\right)  ^{4}}}\\
0 & 1 & 0 & 0\\
0 & 0 & 1 & 0\\
-\frac{1}{\sqrt{2}}\frac{1-\left(  1-\gamma\right)  ^{2}}{\sqrt{1+\left(
1-\gamma\right)  ^{4}}} & 0 & 0 & \frac{1}{\sqrt{2}}\frac{1+\left(
1-\gamma\right)  ^{2}}{\sqrt{1+\left(  1-\gamma\right)  ^{4}}}%
\end{array}
\right)  \text{.} \label{uuu}%
\end{equation}
Finally, the Leung et \textit{al}. recovery operator associated with the
enlarged error operator $A_{0000}$ is given by,%
\begin{equation}
R_{0000}=P_{\mathcal{C}}U_{0000}^{\dagger}\text{,}%
\end{equation}
with $P_{\mathcal{C}}$ in Eq. (\ref{pa}) and $U_{0000}$ in Eq. (\ref{uuu}).

\section{Self-complementary codes}

\subsection{Part 1}

Let $\mathcal{C}$ be a $\left[  \left[  n,k,d\right]  \right]  $ quantum
stabilizer code that spans a $2^{k}$-dimensional subspace of a $2^{n}%
$-dimensional Hilbert space. Two quantum codes $\mathcal{C}^{\left(  1\right)
}$ and $\mathcal{C}^{\left(  2\right)  }$ are \emph{locally permutation
equivalent} if $\mathcal{C}^{\left(  2\right)  }=\tau\mathcal{C}^{\left(
1\right)  }$ with $\tau\overset{\text{def}}{=}\pi T$ where $T$ is a local
unitary transformation in $U\left(  2\right)  ^{\otimes n}$ and $\pi$ is a
permutation of the qubits. When $\mathcal{C}^{\left(  2\right)  }%
=\tau\mathcal{C}^{\left(  1\right)  }$ with $\tau\overset{\text{def}}{=}T$, we
say that the two quantum codes are \emph{locally equivalent}. Finally, we say
that the two codes are globally equivalent, or simply equivalent, if
$\mathcal{C}^{\left(  1\right)  }$ is locally equivalent to a code obtained
from $\mathcal{C}^{\left(  2\right)  }$ by a permutation on qubits.

Assuming single-qubit encoding, how many pairs $\left(  \left\vert
0_{L}\right\rangle \text{, }\left\vert 1_{L}\right\rangle \right)  $
\cite{shor2},
\begin{equation}
\left(  \left\vert 0_{L}\right\rangle \overset{\text{def}}{=}\frac{\left\vert
u\right\rangle +\left\vert \bar{u}\right\rangle }{\sqrt{2}}\text{, }\left\vert
1_{L}\right\rangle \overset{\text{def}}{=}\frac{\left\vert v\right\rangle
+\left\vert \bar{v}\right\rangle }{\sqrt{2}}\right)  \text{,}%
\end{equation}
of orthonormal self-complementary codewords can we construct in the
\emph{complex} Hilbert space $\mathcal{H}_{2}^{4}$? To be explicit, recall
that the canonical computational basis of $\mathcal{H}_{2}^{4}$ reads,%
\begin{equation}
\mathcal{B}_{\mathcal{H}_{2}^{4}}\overset{\text{def}}{=}\left\{
\begin{array}
[c]{c}%
\left\vert e_{0000}\right\rangle \overset{\text{def}}{=}\left\vert
0000\right\rangle \text{,}\left\vert e_{1000}\right\rangle \overset
{\text{def}}{=}\left\vert 1000\right\rangle \text{, }\left\vert e_{0100}%
\right\rangle \overset{\text{def}}{=}\left\vert 0100\right\rangle \text{,
}\left\vert e_{0010}\right\rangle \overset{\text{def}}{=}\left\vert
0010\right\rangle \text{,}\\
\text{ }\left\vert e_{0001}\right\rangle \overset{\text{def}}{=}\left\vert
0001\right\rangle \text{, }\left\vert e_{1100}\right\rangle \overset
{\text{def}}{=}\left\vert 1100\right\rangle \text{,}\left\vert e_{1010}%
\right\rangle \overset{\text{def}}{=}\left\vert 1010\right\rangle \text{,
}\left\vert e_{1001}\right\rangle \overset{\text{def}}{=}\left\vert
1001\right\rangle \text{, }\left\vert e_{0110}\right\rangle \overset
{\text{def}}{=}\left\vert 0110\right\rangle \text{,}\\
\text{ }\left\vert e_{0101}\right\rangle \overset{\text{def}}{=}\left\vert
0101\right\rangle \text{,}\left\vert e_{0011}\right\rangle \overset
{\text{def}}{=}\left\vert 0011\right\rangle \text{,}\left\vert e_{1110}%
\right\rangle \overset{\text{def}}{=}\left\vert 1110\right\rangle \text{,
}\left\vert e_{1101}\right\rangle \overset{\text{def}}{=}\left\vert
1101\right\rangle \text{, }\left\vert e_{0111}\right\rangle \overset
{\text{def}}{=}\left\vert 0111\right\rangle \text{,}\\
\left\vert e_{1011}\right\rangle \overset{\text{def}}{=}\left\vert
1011\right\rangle \text{, }\left\vert e_{1111}\right\rangle \overset
{\text{def}}{=}\left\vert 1111\right\rangle \text{ }%
\end{array}
\right\}  \text{.}%
\end{equation}
The number of possible pairs is,%
\begin{equation}
\#\text{ pairs of orthonormal self-complementary codewords in }\mathcal{H}%
_{2}^{4}=\frac{8^{2}-8}{2}=28\text{, }%
\end{equation}
and they are given by,%
\begin{equation}
\left(  \left\vert v_{i}^{\left(  +\right)  }\right\rangle \text{, }\left\vert
v_{j}^{\left(  +\right)  }\right\rangle \right)  \text{, }i<j\in\left\{
1\text{, }2\text{,..., }8\right\}  \text{,}%
\end{equation}
where,%
\begin{align}
&  \left\vert v_{1}^{\left(  +\right)  }\right\rangle \overset{\text{def}}%
{=}\frac{\left\vert 0000\right\rangle +\left\vert 1111\right\rangle \text{ }%
}{\sqrt{2}}\text{, }\left\vert v_{2}^{\left(  +\right)  }\right\rangle
\overset{\text{def}}{=}\frac{\left\vert 1000\right\rangle +\left\vert
0111\right\rangle \text{ }}{\sqrt{2}}\text{, }\left\vert v_{3}^{\left(
+\right)  }\right\rangle \overset{\text{def}}{=}\frac{\left\vert
0100\right\rangle +\left\vert 1011\right\rangle \text{ }}{\sqrt{2}}\text{,
}\left\vert v_{4}^{\left(  +\right)  }\right\rangle \overset{\text{def}}%
{=}\frac{\left\vert 0010\right\rangle +\left\vert 1101\right\rangle \text{ }%
}{\sqrt{2}}\text{,}\nonumber\\
& \nonumber\\
&  \left\vert v_{5}^{\left(  +\right)  }\right\rangle \overset{\text{def}}%
{=}\frac{\left\vert 0001\right\rangle +\left\vert 1110\right\rangle \text{ }%
}{\sqrt{2}}\text{, }\left\vert v_{6}^{\left(  +\right)  }\right\rangle
\overset{\text{def}}{=}\frac{\left\vert 1100\right\rangle +\left\vert
0011\right\rangle \text{ }}{\sqrt{2}}\text{, }\left\vert v_{7}^{\left(
+\right)  }\right\rangle \overset{\text{def}}{=}\frac{\left\vert
1010\right\rangle +\left\vert 0101\right\rangle \text{ }}{\sqrt{2}}\text{,
}\left\vert v_{8}^{\left(  +\right)  }\right\rangle \overset{\text{def}}%
{=}\frac{\left\vert 1001\right\rangle +\left\vert 0110\right\rangle \text{ }%
}{\sqrt{2}}\text{.}%
\end{align}
Thus, the possible combinations are%
\begin{align}
&  \left(  \left\vert v_{1}^{\left(  +\right)  }\right\rangle \text{,
}\left\vert v_{2}^{\left(  +\right)  }\right\rangle \right)  \text{, }\left(
\left\vert v_{1}^{\left(  +\right)  }\right\rangle \text{, }\left\vert
v_{3}^{\left(  +\right)  }\right\rangle \right)  \text{, }\left(  \left\vert
v_{1}^{\left(  +\right)  }\right\rangle \text{, }\left\vert v_{4}^{\left(
+\right)  }\right\rangle \right)  \text{, }\left(  \left\vert v_{1}^{\left(
+\right)  }\right\rangle \text{, }\left\vert v_{5}^{\left(  +\right)
}\right\rangle \right)  \text{, }\left(  \left\vert v_{1}^{\left(  +\right)
}\right\rangle \text{, }\left\vert v_{6}^{\left(  +\right)  }\right\rangle
\right)  \text{, }\left(  \left\vert v_{1}^{\left(  +\right)  }\right\rangle
\text{, }\left\vert v_{7}^{\left(  +\right)  }\right\rangle \right)  \text{,
}\nonumber\\
& \nonumber\\
&  \left(  \left\vert v_{1}^{\left(  +\right)  }\right\rangle \text{,
}\left\vert v_{8}^{\left(  +\right)  }\right\rangle \right)  \text{, }\left(
\left\vert v_{2}^{\left(  +\right)  }\right\rangle \text{, }\left\vert
v_{3}^{\left(  +\right)  }\right\rangle \right)  \text{, }\left(  \left\vert
v_{2}^{\left(  +\right)  }\right\rangle \text{, }\left\vert v_{4}^{\left(
+\right)  }\right\rangle \right)  \text{, }\left(  \left\vert v_{2}^{\left(
+\right)  }\right\rangle \text{, }\left\vert v_{5}^{\left(  +\right)
}\right\rangle \right)  \text{, }\left(  \left\vert v_{2}^{\left(  +\right)
}\right\rangle \text{, }\left\vert v_{6}^{\left(  +\right)  }\right\rangle
\right)  \text{, }\nonumber\\
& \nonumber\\
&  \left(  \left\vert v_{2}^{\left(  +\right)  }\right\rangle \text{,
}\left\vert v_{7}^{\left(  +\right)  }\right\rangle \right)  \text{, }\left(
\left\vert v_{2}^{\left(  +\right)  }\right\rangle \text{, }\left\vert
v_{8}^{\left(  +\right)  }\right\rangle \right)  \text{, }\left(  \left\vert
v_{3}^{\left(  +\right)  }\right\rangle \text{, }\left\vert v_{4}^{\left(
+\right)  }\right\rangle \right)  \text{, }\left(  \left\vert v_{3}^{\left(
+\right)  }\right\rangle \text{, }\left\vert v_{5}^{\left(  +\right)
}\right\rangle \right)  \text{, }\left(  \left\vert v_{3}^{\left(  +\right)
}\right\rangle \text{, }\left\vert v_{6}^{\left(  +\right)  }\right\rangle
\right)  \text{, }\left(  \left\vert v_{3}^{\left(  +\right)  }\right\rangle
\text{, }\left\vert v_{7}^{\left(  +\right)  }\right\rangle \right)
\text{,}\nonumber\\
& \nonumber\\
&  \left(  \left\vert v_{3}^{\left(  +\right)  }\right\rangle \text{,
}\left\vert v_{8}^{\left(  +\right)  }\right\rangle \right)  \text{, }\left(
\left\vert v_{4}^{\left(  +\right)  }\right\rangle \text{, }\left\vert
v_{5}^{\left(  +\right)  }\right\rangle \right)  \text{, }\left(  \left\vert
v_{4}^{\left(  +\right)  }\right\rangle \text{, }\left\vert v_{6}^{\left(
+\right)  }\right\rangle \right)  \text{, }\left(  \left\vert v_{4}^{\left(
+\right)  }\right\rangle \text{, }\left\vert v_{7}^{\left(  +\right)
}\right\rangle \right)  \text{, }\left(  \left\vert v_{4}^{\left(  +\right)
}\right\rangle \text{, }\left\vert v_{8}^{\left(  +\right)  }\right\rangle
\right)  \text{, }\left(  \left\vert v_{5}^{\left(  +\right)  }\right\rangle
\text{, }\left\vert v_{6}^{\left(  +\right)  }\right\rangle \right)  \text{,
}\nonumber\\
& \nonumber\\
&  \left(  \left\vert v_{5}^{\left(  +\right)  }\right\rangle \text{,
}\left\vert v_{7}^{\left(  +\right)  }\right\rangle \right)  \text{, }\left(
\left\vert v_{5}^{\left(  +\right)  }\right\rangle \text{, }\left\vert
v_{8}^{\left(  +\right)  }\right\rangle \right)  \text{, }\left(  \left\vert
v_{6}^{\left(  +\right)  }\right\rangle \text{, }\left\vert v_{7}^{\left(
+\right)  }\right\rangle \right)  \text{, }\left(  \left\vert v_{6}^{\left(
+\right)  }\right\rangle \text{, }\left\vert v_{8}^{\left(  +\right)
}\right\rangle \right)  \text{, }\left(  \left\vert v_{7}^{\left(  +\right)
}\right\rangle \text{, }\left\vert v_{8}^{\left(  +\right)  }\right\rangle
\right)  \text{.}%
\end{align}
It turns out that among the $\binom{8}{2}=28$-pairs of possible
self-complementary orthogonal codewords in $\mathcal{H}_{2}^{4}$, only three
pairs are indeed good single-AD error correcting codes. For more details, see
Part $2$ of Appendix C. They are given by:

\begin{itemize}
\item The $\left(  \left\vert v_{1}^{\left(  +\right)  }\right\rangle \text{,
}\left\vert v_{6}^{\left(  +\right)  }\right\rangle \right)  $-pair that
represents the Leung et \textit{al}. $\left[  \left[  4\text{, }1\right]
\right]  $-code. The non-normalized codewords read,%
\begin{equation}
\left\vert 0_{L}\right\rangle \overset{\text{def}}{=}\left\vert
0000\right\rangle +\left\vert 1111\right\rangle \text{ and, }\left\vert
1_{L}\right\rangle \overset{\text{def}}{=}\left\vert 0011\right\rangle
+\left\vert 1100\right\rangle \text{.} \label{a}%
\end{equation}

\item The $\left(  \left\vert v_{1}^{\left(  +\right)  }\right\rangle \text{,
}\left\vert v_{8}^{\left(  +\right)  }\right\rangle \right)  $-pair that
represents the Grassl et \textit{al}. perfect quantum erasure code. The
non-normalized codewords read,%
\begin{equation}
\left\vert 0_{L}\right\rangle \overset{\text{def}}{=}\left\vert
0000\right\rangle +\left\vert 1111\right\rangle \text{ and, }\left\vert
1_{L}\right\rangle \overset{\text{def}}{=}\left\vert 1001\right\rangle
+\left\vert 0110\right\rangle \text{.} \label{b}%
\end{equation}

\item The $\left(  \left\vert v_{1}^{\left(  +\right)  }\right\rangle \text{,
}\left\vert v_{7}^{\left(  +\right)  }\right\rangle \right)  $-pair has no
specific mention in the literature, to the best of our knowledge. The
non-normalized codewords read,%
\begin{equation}
\left\vert 0_{L}\right\rangle \overset{\text{def}}{=}\left\vert
0000\right\rangle +\left\vert 1111\right\rangle \text{ and, }\left\vert
1_{L}\right\rangle \overset{\text{def}}{=}\left\vert 0101\right\rangle
+\left\vert 1010\right\rangle \text{.} \label{c}%
\end{equation}

\end{itemize}

The three codes spanned by the codewords in (\ref{a}), (\ref{b}) and (\ref{c})
are indeed globally equivalent. We point out that the codeword $\left\vert
0_{L}\right\rangle \overset{\text{def}}{=}\left\vert 0000\right\rangle
+\left\vert 1111\right\rangle $ is the only codeword in these $28$-pairs that
is invariant under any cyclical permutation of qubits. This property of
$\left\vert 0_{L}\right\rangle $ turns out to be very useful when checking out
the global equivalence among the three good single-AD error correcting codes.
In particular, the Leung et \textit{al}. $\left[  \left[  4\text{, }1\right]
\right]  $-code is globally equivalent to the Grassl et \textit{al.} perfect
quantum erasure code encoding one qubit and correcting one arbitrary erasure.

\subsection{Part 2}

Observe that for the Leung et \textit{al}. four-qubit code (normalization
factors are omitted), we have%
\begin{align}
A_{0000}\left\vert 0_{L}\right\rangle  &  =\left\vert 0000\right\rangle
+\left(  1-\gamma\right)  ^{2}\left\vert 1111\right\rangle \text{, }%
A_{0000}\left\vert 1_{L}\right\rangle =\left(  1-\gamma\right)  \left\vert
0011\right\rangle +\left(  1-\gamma\right)  \left\vert 1100\right\rangle
\text{,}\nonumber\\
A_{1000}\left\vert 0_{L}\right\rangle  &  =\sqrt{\gamma}\left(  1-\gamma
\right)  ^{\frac{3}{2}}\left\vert 0111\right\rangle \text{, }A_{1000}%
\left\vert 1_{L}\right\rangle =\sqrt{\gamma\left(  1-\gamma\right)
}\left\vert 0100\right\rangle \text{,}\nonumber\\
A_{0100}\left\vert 0_{L}\right\rangle  &  =\sqrt{\gamma}\left(  1-\gamma
\right)  ^{\frac{3}{2}}\left\vert 1011\right\rangle \text{, }A_{0100}%
\left\vert 1_{L}\right\rangle =\sqrt{\gamma\left(  1-\gamma\right)
}\left\vert 1000\right\rangle \text{,}\nonumber\\
A_{0010}\left\vert 0_{L}\right\rangle  &  =\sqrt{\gamma}\left(  1-\gamma
\right)  ^{\frac{3}{2}}\left\vert 1101\right\rangle \text{, }A_{0010}%
\left\vert 1_{L}\right\rangle =\sqrt{\gamma\left(  1-\gamma\right)
}\left\vert 0001\right\rangle \text{,}\nonumber\\
A_{0001}\left\vert 0_{L}\right\rangle  &  =\sqrt{\gamma}\left(  1-\gamma
\right)  ^{\frac{3}{2}}\left\vert 1110\right\rangle \text{, }A_{0001}%
\left\vert 1_{L}\right\rangle =\sqrt{\gamma\left(  1-\gamma\right)
}\left\vert 0010\right\rangle \text{.} \label{unos}%
\end{align}
For the Grassl et \textit{al}. four-qubit code (normalization factors are
omitted), we get%
\begin{align}
A_{0000}\left\vert 0_{L}\right\rangle  &  =\left\vert 0000\right\rangle
+\left(  1-\gamma\right)  ^{2}\left\vert 1111\right\rangle \text{, }%
A_{0000}\left\vert 1_{L}\right\rangle =\left(  1-\gamma\right)  \left\vert
1001\right\rangle +\left(  1-\gamma\right)  \left\vert 0110\right\rangle
\text{,}\nonumber\\
A_{1000}\left\vert 0_{L}\right\rangle  &  =\sqrt{\gamma}\left(  1-\gamma
\right)  ^{\frac{3}{2}}\left\vert 0111\right\rangle \text{, }A_{1000}%
\left\vert 1_{L}\right\rangle =\sqrt{\gamma\left(  1-\gamma\right)
}\left\vert 0001\right\rangle \text{,}\nonumber\\
A_{0100}\left\vert 0_{L}\right\rangle  &  =\sqrt{\gamma}\left(  1-\gamma
\right)  ^{\frac{3}{2}}\left\vert 1011\right\rangle \text{, }A_{0100}%
\left\vert 1_{L}\right\rangle =\sqrt{\gamma\left(  1-\gamma\right)
}\left\vert 0010\right\rangle \text{,}\nonumber\\
A_{0010}\left\vert 0_{L}\right\rangle  &  =\sqrt{\gamma}\left(  1-\gamma
\right)  ^{\frac{3}{2}}\left\vert 1101\right\rangle \text{, }A_{0010}%
\left\vert 1_{L}\right\rangle =\sqrt{\gamma\left(  1-\gamma\right)
}\left\vert 0100\right\rangle \text{,}\nonumber\\
A_{0001}\left\vert 0_{L}\right\rangle  &  =\sqrt{\gamma}\left(  1-\gamma
\right)  ^{\frac{3}{2}}\left\vert 1110\right\rangle \text{, }A_{0001}%
\left\vert 1_{L}\right\rangle =\sqrt{\gamma\left(  1-\gamma\right)
}\left\vert 1000\right\rangle \text{.} \label{du}%
\end{align}
Finally, for the third\textbf{ }four-qubit code defined in Eq. (\ref{c})
(normalization factors are omitted), we obtain%
\begin{align}
A_{0000}\left\vert 0_{L}\right\rangle  &  =\left\vert 0000\right\rangle
+\left(  1-\gamma\right)  ^{2}\left\vert 1111\right\rangle \text{, }%
A_{0000}\left\vert 1_{L}\right\rangle =\left(  1-\gamma\right)  \left\vert
0101\right\rangle +\left(  1-\gamma\right)  \left\vert 1010\right\rangle
\text{,}\nonumber\\
A_{1000}\left\vert 0_{L}\right\rangle  &  =\sqrt{\gamma}\left(  1-\gamma
\right)  ^{\frac{3}{2}}\left\vert 0111\right\rangle \text{, }A_{1000}%
\left\vert 1_{L}\right\rangle =\sqrt{\gamma\left(  1-\gamma\right)
}\left\vert 0010\right\rangle \text{,}\nonumber\\
A_{0100}\left\vert 0_{L}\right\rangle  &  =\sqrt{\gamma}\left(  1-\gamma
\right)  ^{\frac{3}{2}}\left\vert 1011\right\rangle \text{, }A_{0100}%
\left\vert 1_{L}\right\rangle =\sqrt{\gamma\left(  1-\gamma\right)
}\left\vert 0001\right\rangle \text{,}\nonumber\\
A_{0010}\left\vert 0_{L}\right\rangle  &  =\sqrt{\gamma}\left(  1-\gamma
\right)  ^{\frac{3}{2}}\left\vert 1101\right\rangle \text{, }A_{0010}%
\left\vert 1_{L}\right\rangle =\sqrt{\gamma\left(  1-\gamma\right)
}\left\vert 1000\right\rangle \text{,}\nonumber\\
A_{0001}\left\vert 0_{L}\right\rangle  &  =\sqrt{\gamma}\left(  1-\gamma
\right)  ^{\frac{3}{2}}\left\vert 1110\right\rangle \text{, }A_{0001}%
\left\vert 1_{L}\right\rangle =\sqrt{\gamma\left(  1-\gamma\right)
}\left\vert 0100\right\rangle \text{.} \label{tres}%
\end{align}
From Eqs. (\ref{unos}), (\ref{du}) and (\ref{tres}) it is straightforward to
show that the pairs $\left(  \left\vert v_{1}^{\left(  +\right)
}\right\rangle \text{, }\left\vert v_{6}^{\left(  +\right)  }\right\rangle
\right)  $, $\left(  \left\vert v_{1}^{\left(  +\right)  }\right\rangle
\text{, }\left\vert v_{8}^{\left(  +\right)  }\right\rangle \right)  $ and
$\left(  \left\vert v_{1}^{\left(  +\right)  }\right\rangle \text{,
}\left\vert v_{7}^{\left(  +\right)  }\right\rangle \right)  $, respectively,
lead to good codes for the AD errors. Finally, it can be checked that,%
\begin{align}
&  \left(  \left\vert v_{1}^{\left(  +\right)  }\right\rangle \text{,
}\left\vert v_{2}^{\left(  +\right)  }\right\rangle \right)  \text{ is not
good because }\left\{  A_{0000}\text{, }A_{1000}\right\}  \text{ is not
correctable; }\nonumber\\
&  \left(  \left\vert v_{1}^{\left(  +\right)  }\right\rangle \text{,
}\left\vert v_{3}^{\left(  +\right)  }\right\rangle \right)  \text{, is not
good because }\left\{  A_{0000}\text{, }A_{0100}\right\}  \text{ is not
correctable;}\nonumber\\
&  \left(  \left\vert v_{1}^{\left(  +\right)  }\right\rangle \text{,
}\left\vert v_{4}^{\left(  +\right)  }\right\rangle \right)  \text{, is not
good because }\left\{  A_{0000}\text{, }A_{0010}\right\}  \text{ is not
correctable;}\nonumber\\
&  \left(  \left\vert v_{1}^{\left(  +\right)  }\right\rangle \text{,
}\left\vert v_{5}^{\left(  +\right)  }\right\rangle \right)  \text{, is not
good because }\left\{  A_{0000}\text{, }A_{0001}\right\}  \text{ is not
correctable;}\nonumber\\
&  \left(  \left\vert v_{2}^{\left(  +\right)  }\right\rangle \text{,
}\left\vert v_{3}^{\left(  +\right)  }\right\rangle \right)  \text{, is not
good because }\left\{  A_{1000}\text{, }A_{0100}\right\}  \text{ is not
correctable;}\nonumber\\
&  \left(  \left\vert v_{2}^{\left(  +\right)  }\right\rangle \text{,
}\left\vert v_{4}^{\left(  +\right)  }\right\rangle \right)  \text{, is not
good because }\left\{  A_{1000}\text{, }A_{0010}\right\}  \text{ is not
correctable;}\nonumber\\
&  \left(  \left\vert v_{2}^{\left(  +\right)  }\right\rangle \text{,
}\left\vert v_{5}^{\left(  +\right)  }\right\rangle \right)  \text{, is not
good because }\left\{  A_{1000}\text{, }A_{0001}\right\}  \text{ is not
correctable;}\nonumber\\
&  \left(  \left\vert v_{2}^{\left(  +\right)  }\right\rangle \text{,
}\left\vert v_{6}^{\left(  +\right)  }\right\rangle \right)  \text{, is not
good because }\left\{  A_{0000}\text{, }A_{0100}\right\}  \text{ is not
correctable;}\nonumber\\
&  \left(  \left\vert v_{2}^{\left(  +\right)  }\right\rangle \text{,
}\left\vert v_{7}^{\left(  +\right)  }\right\rangle \right)  \text{, is not
good because }\left\{  A_{0000}\text{, }A_{0010}\right\}  \text{ is not
correctable;}\nonumber\\
&  \left(  \left\vert v_{2}^{\left(  +\right)  }\right\rangle \text{,
}\left\vert v_{8}^{\left(  +\right)  }\right\rangle \right)  \text{, is not
good because }\left\{  A_{0000}\text{, }A_{0001}\right\}  \text{ is not
correctable,}%
\end{align}
and,%
\begin{align}
&  \left(  \left\vert v_{3}^{\left(  +\right)  }\right\rangle \text{,
}\left\vert v_{4}^{\left(  +\right)  }\right\rangle \right)  \text{, is not
good because }\left\{  A_{0100}\text{, }A_{0010}\right\}  \text{ is not
correctable;}\nonumber\\
&  \left(  \left\vert v_{3}^{\left(  +\right)  }\right\rangle \text{,
}\left\vert v_{5}^{\left(  +\right)  }\right\rangle \right)  \text{, is not
good because }\left\{  A_{0100}\text{, }A_{0001}\right\}  \text{ is not
correctable;}\nonumber\\
&  \left(  \left\vert v_{3}^{\left(  +\right)  }\right\rangle \text{,
}\left\vert v_{6}^{\left(  +\right)  }\right\rangle \right)  \text{, is not
good because }\left\{  A_{0000}\text{, }A_{1000}\right\}  \text{ is not
correctable;}\nonumber\\
&  \left(  \left\vert v_{3}^{\left(  +\right)  }\right\rangle \text{,
}\left\vert v_{7}^{\left(  +\right)  }\right\rangle \right)  \text{, is not
good because }\left\{  A_{0000}\text{, }A_{0001}\right\}  \text{ is not
correctable;}\nonumber\\
&  \left(  \left\vert v_{3}^{\left(  +\right)  }\right\rangle \text{,
}\left\vert v_{8}^{\left(  +\right)  }\right\rangle \right)  \text{, is not
good because }\left\{  A_{0000}\text{, }A_{0010}\right\}  \text{ is not
correctable;}\nonumber\\
&  \left(  \left\vert v_{4}^{\left(  +\right)  }\right\rangle \text{,
}\left\vert v_{5}^{\left(  +\right)  }\right\rangle \right)  \text{, is not
good because }\left\{  A_{0010}\text{, }A_{0001}\right\}  \text{ is not
correctable;}\nonumber\\
&  \left(  \left\vert v_{4}^{\left(  +\right)  }\right\rangle \text{,
}\left\vert v_{6}^{\left(  +\right)  }\right\rangle \right)  \text{, is not
good because }\left\{  A_{0000}\text{, }A_{0001}\right\}  \text{ is not
correctable;}\nonumber\\
&  \left(  \left\vert v_{4}^{\left(  +\right)  }\right\rangle \text{,
}\left\vert v_{7}^{\left(  +\right)  }\right\rangle \right)  \text{, is not
good because }\left\{  A_{0000}\text{, }A_{1000}\right\}  \text{ is not
correctable;}\nonumber\\
&  \left(  \left\vert v_{4}^{\left(  +\right)  }\right\rangle \text{,
}\left\vert v_{8}^{\left(  +\right)  }\right\rangle \right)  \text{, is not
good because }\left\{  A_{0000}\text{, }A_{0100}\right\}  \text{ is not
correctable.}%
\end{align}
Finally, we have that%
\begin{align}
&  \left(  \left\vert v_{5}^{\left(  +\right)  }\right\rangle \text{,
}\left\vert v_{6}^{\left(  +\right)  }\right\rangle \right)  \text{, is not
good because }\left\{  A_{0000}\text{, }A_{0010}\right\}  \text{ is not
correctable;}\nonumber\\
&  \left(  \left\vert v_{5}^{\left(  +\right)  }\right\rangle \text{,
}\left\vert v_{7}^{\left(  +\right)  }\right\rangle \right)  \text{, is not
good because }\left\{  A_{0000}\text{, }A_{0100}\right\}  \text{ is not
correctable;}\nonumber\\
&  \left(  \left\vert v_{5}^{\left(  +\right)  }\right\rangle \text{,
}\left\vert v_{8}^{\left(  +\right)  }\right\rangle \right)  \text{, is not
good because }\left\{  A_{0000}\text{, }A_{1000}\right\}  \text{ is not
correctable;}\nonumber\\
&  \left(  \left\vert v_{6}^{\left(  +\right)  }\right\rangle \text{,
}\left\vert v_{7}^{\left(  +\right)  }\right\rangle \right)  \text{, is not
good because }\left\{  A_{0100}\text{, }A_{0010}\right\}  \text{ is not
correctable;}\nonumber\\
&  \left(  \left\vert v_{6}^{\left(  +\right)  }\right\rangle \text{,
}\left\vert v_{8}^{\left(  +\right)  }\right\rangle \right)  \text{, is not
good because }\left\{  A_{0100}\text{, }A_{0001}\right\}  \text{ is not
correctable;}\nonumber\\
&  \left(  \left\vert v_{7}^{\left(  +\right)  }\right\rangle \text{,
}\left\vert v_{8}^{\left(  +\right)  }\right\rangle \right)  \text{, is not
good because }\left\{  A_{0010}\text{, }A_{0001}\right\}  \text{ is not
correctable.}%
\end{align}

\section{The complex optimization problem}

In this Appendix, we use the notation $a\equiv\alpha$ and $b\equiv\beta$. The
problem is to find $\bar{\alpha}$ and $\bar{\beta}$ (perhaps $\gamma
$-dependent quantities) such that $\mathcal{F}\left(  \bar{\alpha}\text{,
}\bar{\beta}\text{, }\gamma\right)  $ denotes the searched maximum,%
\begin{equation}
\mathcal{F}\left(  \bar{\alpha}\text{, }\bar{\beta}\text{, }\gamma\right)
=\underset{\left\vert \alpha\right\vert ^{2}+\left\vert \beta\right\vert
^{2}=1}{\max}\mathcal{F}\left(  \alpha\text{, }\beta\text{, }\gamma\right)
\text{.}%
\end{equation}
Let us precede by brute force in an analytical fashion. Assume that,%
\begin{equation}
\alpha=\operatorname{Re}\alpha+i\operatorname{Im}\alpha=\alpha_{R}+i\alpha
_{I}\text{ and, }\beta=\operatorname{Re}\beta+i\operatorname{Im}\beta
=\beta_{R}+i\beta_{I}\text{.}%
\end{equation}
Therefore, the two \emph{complex}-variables complex optimization problem may
be defined in terms of four \emph{real}-variables optimization problem,%
\begin{equation}
\mathcal{F}\left(  \bar{\alpha}_{R}\text{, }\bar{\alpha}_{I}\text{, }%
\bar{\beta}_{R}\text{, }\bar{\beta}_{I}\text{, }\gamma\right)  =\underset
{\bar{\alpha}_{R}^{2}+\bar{\alpha}_{I}^{2}+\bar{\beta}_{R}^{2}+\bar{\beta}%
_{I}^{2}=1}{\max}\mathcal{F}\left(  \alpha_{R}\text{, }\alpha_{I}\text{,
}\beta_{R}\text{, }\beta_{I}\text{, }\gamma\right)  \text{.}%
\end{equation}
Observe that $\mathcal{F}_{\left[  \left[  4,1\right]  \right]  }\left(
\alpha\text{, }\beta\text{, }\gamma\right)  $ can be rewritten as,
\begin{equation}
\mathcal{F}_{\left[  \left[  4,1\right]  \right]  }\left(  \gamma\right)
\left(  \alpha\text{, }\beta\text{, }\gamma\right)  =\frac{1}{4}\left\{
A+B+2\gamma\left(  1-\gamma\right)  \left(  2-\gamma\right)  ^{2}+2\gamma
^{2}\left(  1-\gamma\right)  ^{2}+\frac{\gamma^{4}}{2}\right\}  \text{,}
\label{amo}%
\end{equation}
where,%
\begin{align}
A\overset{\text{def}}{=}\left\vert \frac{\alpha+\beta\left(  1-\gamma\right)
^{2}}{\sqrt{2}}+\left(  1-\gamma\right)  \right\vert ^{2}  &  =\left\vert
\frac{\left(  \alpha_{R}+i\alpha_{I}\right)  +\left(  \beta_{R}+i\beta
_{I}\right)  \left(  1-\gamma\right)  ^{2}}{\sqrt{2}}+\left(  1-\gamma\right)
\right\vert ^{2}\nonumber\\
& \nonumber\\
&  =\left(  \frac{\alpha_{R}+\beta_{R}\left(  1-\gamma\right)  ^{2}}{\sqrt{2}%
}+\left(  1-\gamma\right)  \right)  ^{2}+\left(  \frac{\alpha_{I}+\beta
_{I}\left(  1-\gamma\right)  ^{2}}{\sqrt{2}}\right)  ^{2}\text{,} \label{x}%
\end{align}
and,%
\begin{align}
B\overset{\text{def}}{=}\left\vert \frac{\beta^{\ast}-\alpha^{\ast}\left(
1-\gamma\right)  ^{2}}{\sqrt{2}}\right\vert ^{2}  &  =\left\vert \frac{\left(
\beta_{R}-i\beta_{I}\right)  -\left(  \alpha_{R}-i\alpha_{I}\right)  \left(
1-\gamma\right)  ^{2}}{\sqrt{2}}\right\vert ^{2}=\left\vert \frac{\beta
_{R}-\alpha_{R}\left(  1-\gamma\right)  ^{2}}{\sqrt{2}}-i\frac{\beta
_{I}-\alpha_{I}\left(  1-\gamma\right)  ^{2}}{\sqrt{2}}\right\vert
^{2}\nonumber\\
& \nonumber\\
&  =\left(  \frac{\beta_{R}-\alpha_{R}\left(  1-\gamma\right)  ^{2}}{\sqrt{2}%
}\right)  ^{2}+\left(  \frac{\beta_{I}-\alpha_{I}\left(  1-\gamma\right)
^{2}}{\sqrt{2}}\right)  ^{2}\text{.} \label{xx}%
\end{align}
After some algebraic manipulation of Eqs. (\ref{xx})\ and (\ref{x}), we get%
\begin{align}
A+B  &  =\left(  \frac{\alpha_{R}+\beta_{R}\left(  1-\gamma\right)  ^{2}%
}{\sqrt{2}}+\left(  1-\gamma\right)  \right)  ^{2}+\left(  \frac{\alpha
_{I}+\beta_{I}\left(  1-\gamma\right)  ^{2}}{\sqrt{2}}\right)  ^{2}+\left(
\frac{\beta_{R}-\alpha_{R}\left(  1-\gamma\right)  ^{2}}{\sqrt{2}}\right)
^{2}+\left(  \frac{\beta_{I}-\alpha_{I}\left(  1-\gamma\right)  ^{2}}{\sqrt
{2}}\right)  ^{2}\nonumber\\
& \nonumber\\
&  =\left[  \left(  \frac{\alpha_{R}+\beta_{R}\left(  1-\gamma\right)  ^{2}%
}{\sqrt{2}}+\left(  1-\gamma\right)  \right)  ^{2}+\left(  \frac{\beta
_{R}-\alpha_{R}\left(  1-\gamma\right)  ^{2}}{\sqrt{2}}\right)  ^{2}\right]
+\nonumber\\
& \nonumber\\
&  +\left[  \left(  \frac{\alpha_{I}+\beta_{I}\left(  1-\gamma\right)  ^{2}%
}{\sqrt{2}}\right)  ^{2}+\left(  \frac{\beta_{I}-\alpha_{I}\left(
1-\gamma\right)  ^{2}}{\sqrt{2}}\right)  ^{2}\right] \nonumber\\
& \nonumber\\
&  =\left[  \left(  \alpha_{R}^{2}+\beta_{R}^{2}\right)  \left(
\frac{1+\left(  1-\gamma\right)  ^{4}}{2}\right)  +\left(  1-\gamma\right)
^{2}+2\left(  1-\gamma\right)  \frac{\alpha_{R}+\beta_{R}\left(
1-\gamma\right)  ^{2}}{\sqrt{2}}\right]  +\nonumber\\
& \nonumber\\
&  +\left[  \left(  \alpha_{I}^{2}+\beta_{I}^{2}\right)  \left(
\frac{1+\left(  1-\gamma\right)  ^{4}}{2}\right)  \right] \nonumber\\
& \nonumber\\
&  =\frac{1+\left(  1-\gamma\right)  ^{4}}{2}+\left(  1-\gamma\right)
^{2}+2\left(  1-\gamma\right)  \frac{\alpha_{R}+\beta_{R}\left(
1-\gamma\right)  ^{2}}{\sqrt{2}}\text{.} \label{chistu}%
\end{align}
Therefore, substituting (\ref{chistu}) into (\ref{amo}), $\mathcal{F}_{\left[
\left[  4,1\right]  \right]  }\left(  \alpha\text{, }\beta\text{, }%
\gamma\right)  $ becomes%
\begin{align}
\mathcal{F}_{\left[  \left[  4,1\right]  \right]  }\left(  \alpha_{R}\text{,
}\alpha_{I}\text{, }\beta_{R}\text{, }\beta_{I}\text{, }\gamma\right)   &
=\frac{1}{4}\left\{
\begin{array}
[c]{c}%
\frac{1+\left(  1-\gamma\right)  ^{4}}{2}+\left(  1-\gamma\right)
^{2}+2\left(  1-\gamma\right)  \frac{\alpha_{R}+\beta_{R}\left(
1-\gamma\right)  ^{2}}{\sqrt{2}}+2\gamma\left(  1-\gamma\right)  \left(
2-\gamma\right)  ^{2}+\\
\\
+2\gamma^{2}\left(  1-\gamma\right)  ^{2}+\frac{\gamma^{4}}{2}%
\end{array}
\right\} \nonumber\\
& \nonumber\\
&  =\mathcal{F}_{0}\left(  \gamma\right)  +\frac{2\alpha_{R}\left(
1-\gamma\right)  +2\beta_{R}\left(  1-\gamma\right)  ^{3}}{4\sqrt{2}}\text{,}%
\end{align}
where,%
\begin{equation}
\mathcal{F}_{0}\left(  \gamma\right)  \overset{\text{def}}{=}\frac{1}%
{4}\left(  \frac{1+\left(  1-\gamma\right)  ^{4}}{2}+\left(  1-\gamma\right)
^{2}+2\gamma\left(  1-\gamma\right)  \left(  2-\gamma\right)  ^{2}+2\gamma
^{2}\left(  1-\gamma\right)  ^{2}+\frac{\gamma^{4}}{2}\right)  \text{.}%
\end{equation}
Therefore, the \emph{complex} optimization problem becomes%
\begin{align}
\mathcal{F}_{\left[  \left[  4,1\right]  \right]  }\left(  \bar{\alpha}%
_{R}\text{, }\bar{\alpha}_{I}\text{, }\bar{\beta}_{R}\text{, }\bar{\beta}%
_{I}\text{, }\gamma\right)   &  =\underset{\bar{\alpha}_{R}^{2}+\bar{\alpha
}_{I}^{2}+\bar{\beta}_{R}^{2}+\bar{\beta}_{I}^{2}=1}{\max}\mathcal{F}_{\left[
\left[  4,1\right]  \right]  }\left(  \alpha_{R}\text{, }\alpha_{I}\text{,
}\beta_{R}\text{, }\beta_{I}\text{, }\gamma\right) \nonumber\\
& \nonumber\\
&  =\underset{\bar{\alpha}_{R}^{2}+\bar{\alpha}_{I}^{2}+\bar{\beta}_{R}%
^{2}+\bar{\beta}_{I}^{2}=1}{\max}\left[  \mathcal{F}_{0}\left(  \gamma\right)
+\frac{2\alpha_{R}\left(  1-\gamma\right)  +2\beta_{R}\left(  1-\gamma\right)
^{3}}{4\sqrt{2}}\right]  \text{.}%
\end{align}
We note that $\mathcal{F}_{\left[  \left[  4,1\right]  \right]  }\left(
\alpha_{R}\text{, }\alpha_{I}\text{, }\beta_{R}\text{, }\beta_{I}\text{,
}\gamma\right)  $ does not depend on $\alpha_{I}$ and $\beta_{I}$. Setting
$\alpha_{I}$ $=$ $\beta_{I}=0$, the maximization problem becomes%
\begin{equation}
\mathcal{F}_{\left[  \left[  4,1\right]  \right]  }\left(  \bar{\alpha}%
_{R}\text{, }\bar{\beta}_{R}\text{, }\gamma\right)  =\underset{\bar{\alpha
}_{R}^{2}+\bar{\beta}_{R}^{2}=1}{\max}\left[  \mathcal{F}_{0}\left(
\gamma\right)  +\frac{2\alpha_{R}\left(  1-\gamma\right)  +2\beta_{R}\left(
1-\gamma\right)  ^{3}}{4\sqrt{2}}\right]  \text{.}%
\end{equation}
We observe that,%
\begin{equation}
\frac{d}{da_{R}}\left(  \mathcal{F}_{0}\left(  \gamma\right)  +\frac
{2\alpha_{R}\left(  1-\gamma\right)  +2\left(  1-a_{R}^{2}\right)  ^{\frac
{1}{2}}\left(  1-\gamma\right)  ^{3}}{4\sqrt{2}}\right)  =0\text{,}%
\end{equation}
implies that,%
\begin{equation}
\alpha_{R}-2\alpha_{R}\gamma-\sqrt{1-\alpha_{R}^{2}}+\alpha_{R}\gamma
^{2}=0\text{,}%
\end{equation}
that is,%
\begin{equation}
\bar{\alpha}_{R}\left(  \gamma\right)  \overset{\text{def}}{=}\frac{1}%
{\sqrt{1+\left(  1-\gamma\right)  ^{4}}}\text{ and, }\bar{\beta}_{R}\left(
\gamma\right)  \overset{\text{def}}{=}\frac{\left(  1-\gamma\right)  ^{2}%
}{\sqrt{1+\left(  1-\gamma\right)  ^{4}}}\text{.}%
\end{equation}
Finally, we obtain%
\begin{equation}
\mathcal{F}_{\left[  \left[  4,1\right]  \right]  }\left(  \bar{\alpha}%
_{R}\text{, }\bar{\beta}_{R}\text{, }\gamma\right)  \overset{\gamma\ll
1}{\approx}1-\frac{3}{2}\gamma^{2}+\mathcal{O}\left(  \gamma^{3}\right)
\text{.}%
\end{equation}
More precisely, we should set $\alpha_{R}^{2}+\beta_{R}^{2}\leq1$ or,
$\alpha_{R}^{2}+\beta_{R}^{2}=r^{2}$ with $r\leq1$. In this case we have,%
\begin{equation}
\frac{d}{d\alpha_{R}}\left(  \frac{2\alpha_{R}\left(  1-\gamma\right)
+2\left(  r^{2}-\alpha_{R}^{2}\right)  ^{\frac{1}{2}}\left(  1-\gamma\right)
^{3}}{4\sqrt{2}}\right)  =\frac{1}{4}\sqrt{2}\frac{\gamma-1}{\sqrt
{r^{2}-\alpha_{R}^{2}}}\left(  \alpha_{R}-\sqrt{r^{2}-\alpha_{R}^{2}}%
-2\gamma\alpha_{R}+\gamma^{2}\alpha_{R}\right)  =0\text{,}%
\end{equation}
that is,
\begin{subequations}
\begin{equation}
\bar{\alpha}_{R}\left(  \gamma\right)  \overset{\text{def}}{=}\frac{r}%
{\sqrt{1+\left(  1-\gamma\right)  ^{4}}}\text{ and, }\bar{\beta}_{R}\left(
\gamma\right)  \overset{\text{def}}{=}\frac{r\left(  1-\gamma\right)  ^{2}%
}{\sqrt{1+\left(  1-\gamma\right)  ^{4}}}\text{.} \tag{C16}%
\end{equation}
Observe that,
\end{subequations}
\begin{equation}
\mathcal{F}_{\left[  \left[  4,1\right]  \right]  }\left(  \bar{\alpha}%
_{R}\text{, }\bar{\beta}_{R}\text{, }\gamma\right)  =\underset{\bar{\alpha
}_{R}^{2}+\bar{\beta}_{R}^{2}=r^{2}}{\max}\left[  \mathcal{F}_{0}\left(
\gamma\right)  +\frac{2\alpha_{R}\left(  1-\gamma\right)  +2\beta_{R}\left(
1-\gamma\right)  ^{3}}{4\sqrt{2}}\right]  \text{,}%
\end{equation}
that is,%
\begin{equation}
\frac{2\alpha_{R}\left(  1-\gamma\right)  +2\beta_{R}\left(  1-\gamma\right)
^{3}}{4\sqrt{2}}=\frac{2\left(  \frac{r}{\sqrt{1+\left(  1-\gamma\right)
^{4}}}\right)  \left(  1-\gamma\right)  +2\left(  \frac{r\left(
1-\gamma\right)  ^{2}}{\sqrt{1+\left(  1-\gamma\right)  ^{4}}}\right)  \left(
1-\gamma\right)  ^{3}}{4\sqrt{2}}\approx\frac{1}{2}r-r\gamma+r\gamma
^{2}+\mathcal{O}\left(  \gamma^{3}\right)  \text{,}%
\end{equation}
thus,%
\begin{equation}
\mathcal{F}_{\left[  \left[  4,1\right]  \right]  }\left(  \bar{\alpha}%
_{R}\text{, }\bar{\beta}_{R}\text{, }\gamma\right)  \approx\frac{1}{2}\left(
1+r\right)  +\left(  1-r\right)  \gamma-\left(  \frac{5}{2}-r^{2}\right)
\gamma^{2}+\mathcal{O}\left(  \gamma^{3}\right)  \text{.}%
\end{equation}
It then turns out that for $r=1$ we obtain the optimal fidelity,%
\begin{equation}
\mathcal{F}_{\left[  \left[  4,1\right]  \right]  }\left(  \bar{\alpha}%
_{R}\text{, }\bar{\beta}_{R}\text{, }\gamma\right)  \approx1-\frac{3}{2}%
\gamma^{2}+\mathcal{O}\left(  \gamma^{3}\right)  \text{.}%
\end{equation}
In conclusion, setting $\alpha_{I}$ $=$ $\beta_{I}=0$ and $\alpha_{R}$,
$\beta_{R}$ given in Eq. (C16), $\mathcal{F}_{\left[  \left[  4,1\right]
\right]  }\left(  \alpha\text{, }\beta\text{, }\gamma\right)  $ in (\ref{amo})
becomes
\begin{equation}
\mathcal{F}_{\left[  \left[  4,1\right]  \right]  }^{\text{F-recovery}}\left(
\gamma\right)  \approx1-\frac{3}{2}\gamma^{2}+\mathcal{O}\left(  \gamma
^{3}\right)  \text{.} \label{finale}%
\end{equation}
The derivation of Eq. (\ref{finale}) concludes our optimization problem.

We emphasize that after completing this work, we have become aware that Eq.
(\ref{finale}) has also appeared in \cite{china}. However, the derivation
presented in \cite{china} is by no means as explicit as the one provided in
our work.

\bigskip

\bigskip

\bigskip
\end{document}